%% file: paper.tex
\documentclass[letter,prd,aps,twocolumn,floatfix,superscriptaddress,nofootinbib]{revtex4}
 
\usepackage{lipsum}
\usepackage{amssymb,amsmath,graphicx}
\usepackage{hyperref}
\usepackage[usenames,dvipsnames]{color}
\usepackage{array}
\usepackage{footnote}

\input{aas_macros.tex}

\begin{document}

\date{\today}
\title{Effects of High Density Phase Transitions on Neutron Star Dynamics}

\author{Steven L. Liebling}
\affiliation{Long Island University, Brookville, New York 11548, USA}

\author{Carlos Palenzuela}
\affiliation{Departament  de  F\'{\i}sica $\&$ IAC3,  Universitat  de  les  Illes  Balears,  Palma  de Mallorca,  Baleares  E-07122,  Spain}

\author{Luis Lehner}
\affiliation{Perimeter Institute, 31 Caroline St, Waterloo, ON N2L 2Y5, Canada}


\begin{abstract}
Various theoretical arguments motivate an expectation of a 
phase transition in matter at extreme densities above nuclear
density,
accompanied by hopes that gravitational wave observations may
reveal the properties of such a transition. Instead of adopting
a particular theory, we consider here a generic
form of first order phase transition using a piecewise polytropic
equation of state, and evolve both isolated neutron stars and
neutron star binaries, including unequal mass binaries and, in some cases,
magnetic field,
 looking at dynamical effects. Of particular
interest are effects that may be observable either via gravitational waves or electromagnetic observations.
\end{abstract}

\maketitle

\section{Introduction}\label{introduction}

Neutron stars are a natural laboratory for matter at extreme densities
with their central densities extending to multiples of nuclear density
($\rho_\mathrm{nuc} = 2 \times 10^{14} \mathrm{g}/\mathrm{cm}^3$).
As such, observations in the gravitational wave~(GW) or electromagnetic wave~(EM) bands
enable a better understanding of the equation of state~(EoS) of
matter at such high densities. The science obtained from
GW observations of the GW170817~\cite{Abbott:2017D} and GW190425~\cite{Abbott:2020uma} binary neutron star mergers by the LIGO/Virgo
collaboration along with EM followup of the former have revealed many aspects of NSs, and
results by NICER in the X-ray band have provided important estimates for both the mass and
radius of a particular NS (PSR J0030+0451)~\cite{Miller:2019cac,Riley:2019yda,Bogdanov_2019,Bilous_2019}. 

NSs probe a region of the EoS not well understood. In particular, 
low densities can be studied in the lab with hadronic matter
and extended to higher densities with theoretical
extrapolations. At the highest densities, perturbative quantum chromodynamics~(pQCD) predicting free quark matter is well accepted. However, between these two regimes there is 
significant uncertainty and, on fairly general terms, reason to expect (at least one) phase transition~(PT). Binary
neutron star mergers are of particular relevance to explore this region. 

Significant work has studied how such a PT inside a NS might manifest and impact potential observables.
In particular, a number of groups have studied numerically the merger of
two NSs described by various EoSs with PT. Early work compared the
DD2 EoS with a PT to a hyperon phase~\cite{Radice:2016rys}.
Another EoS was studied with a conformally flat, SPH code and found differences
in the postmerger oscillation frequencies associated with a 
PT~\cite{Bauswein:2018bma,Bauswein:2019skm,PhysRevD.102.123023}. A similar result with
higher postmerger frequencies were found with a PT to quark matter using
a fully nonlinear, GR HD code~\cite{Most:2018eaw,Most:2019onn,Weih:2019xvw}.
Yet another group studied BNS mergers within an EoS derived from holography,
finding lower oscillation frequencies of post-merger remnants~\cite{Ecker:2019xrw}.

These numerical investigations suggest that more precision in the postmerger
regime is needed with GW observatories to observe these PTs, although 
Ref.~\cite{Chen:2019rja} argues that
tens of current detections may allow current technology to discern PTs.
In contrast though, Ref.~\cite{Annala:2019puf} makes an interesting
argument with current observations that
high mass NSs likely contain a sizable quark matter core.

In this work, with the goal of further identifying potentially relevant features associated
with a PT in NSs, we consider one particular EoS in piecewise polytropic form
to which we add PTs of different forms. 
These PTs are not motivated by some particular theory, but
instead take a generic approach, subject  to mass constraints. We employ
them only within regimes in which they remain causal\footnote{Note that many oft-used EoSs extend into acausal regimes at high densities.}. We then study the dynamics of single
stars (which can be rotating and/or magnetized) and the
mergers of quasi-circular stellar binaries.
This study adds to the body of work  on the
impacts of PT in BNS --in good agreement with the main qualitative features already observed-- as well as considering some novel
aspects. In particular, the potential role of magnetic field, increased accretion, and transitions into and out of
the PT.

\section{Setup}\label{details}

We use the distributed adaptive mesh refinement code {\sc MHDuet}  to evolve the CCZ4 formalism of the Einstein equations coupled to a magnetized perfect fluid, as described in detail in Ref.~\cite{Liebling:2020jlq}. The code is generated with the platform {\it Simflowny}~\cite{arbona13,arbona18} to run under the {\tt SAMRAI} infrastructure \cite{hornung02,gunney16}, which allows excellent parallel scaling over thousands of processors.
The code has been extensively tested, demonstrating the expected convergence rate of the solutions in different GR and MHD scenarios~\cite{Palenzuela:2018sly,2020PhRvD.101l3019V,Liebling:2020jlq}. {\sc MHDuet} employs the Method of Lines,
with a fourth-order Runge-Kutta time integrator which ensures the stability and convergence of the solution for time steps smaller than a factor of $0.4$ times   the grid spacing~\cite{Butcher:2008}. The space-time evolution equations are discretized in space using centered finite-difference, fourth-order-accurate operators for the derivatives, and sixth-order Kreiss-Oliger dissipation to filter
the high-frequency modes unresolved in our grids.  For the fluid, we employ High-Resolution Shock-Capturing (HRSC) methods~\cite{toro97} to deal with the possible appearance of shocks and to take advantage of the existence of weak solutions in the equations. The fluxes at the cell interfaces are calculated by combining the Lax-Friedrich flux splitting formula~\cite{shu98} with the fifth-order, monotonicity-preserving reconstruction method MP5~\cite{suresh97}. 

\begin{figure}
	\includegraphics[width=3.0in]{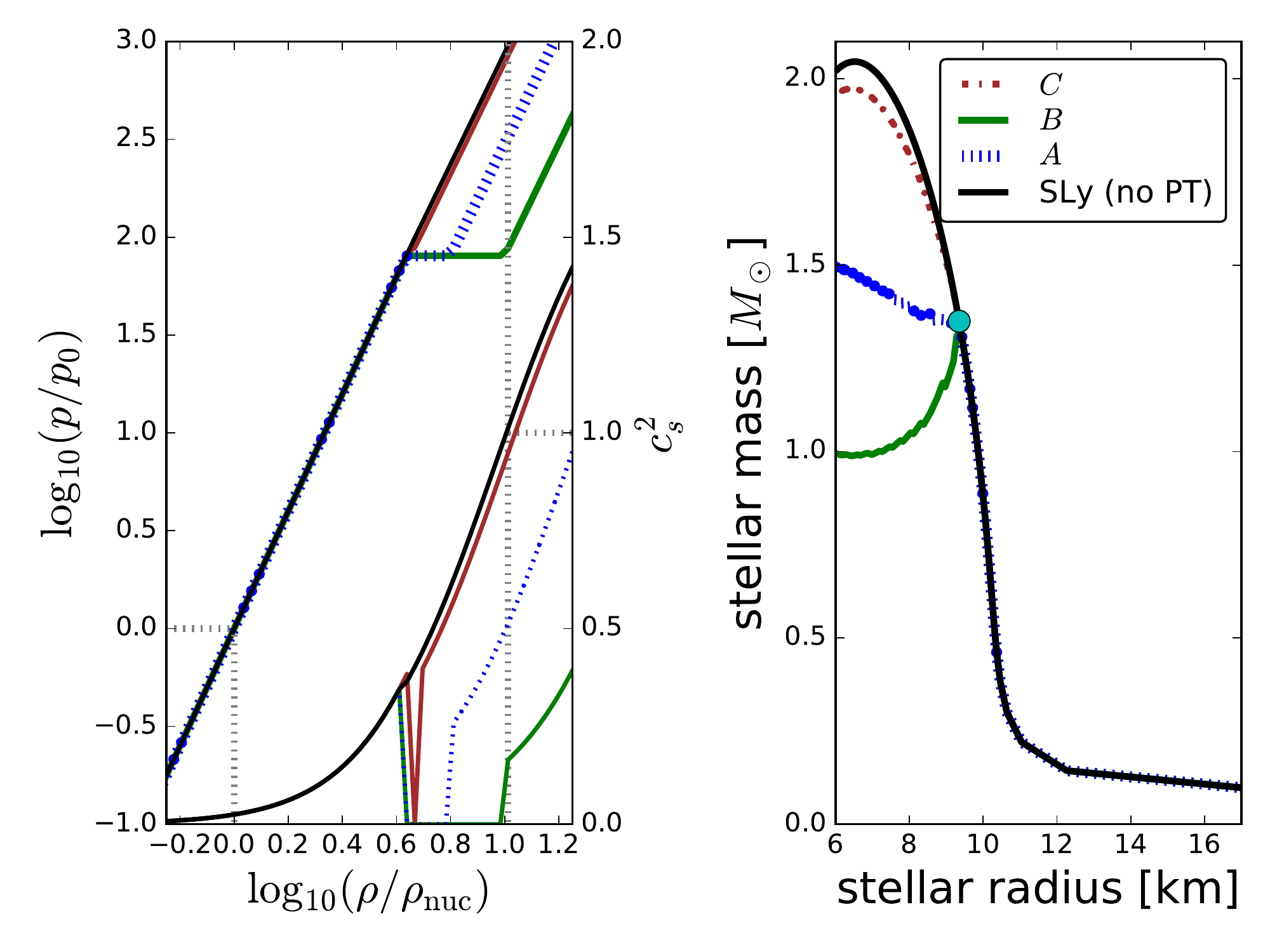}
	\caption{Characterization of the EoSs studied here.
    \textbf{Left:} Standard plot of the pressure versus density
       for the variations of the base SLy EoS. The pressure $p_0$
       is the pressure at nuclear density as shown with gray dotted lines.
       The curve for EoS~C is difficult to distinguish from SLy 
       because its $\Delta_\rho$ is so small. 
       Also shown is the sound speed squared (calculated as in Eq.~\ref{eq:csq})
       for the different EoSs 
        (these are the curves in the bottom right of the panel, contrasting
        with line segments representing the pressure). 
       Note that only those with a `strong' PT
       satisfy $c_s^2\le 1$, an argument that figures into the recent Ref.~\cite{Annala:2019puf}.
    \textbf{Right:} Family of isolated NS solutions (TOV stars)
       corresponding to each EoS showing the (gravitational) mass versus stellar radius.  The cyan circle indicates the
       particular star evolved in Fig.~\ref{fig:singlestar}.
	}
	\label{fig:eoses}
\end{figure}

In our previous study of neutron star mergers~\cite{Liebling:2020jlq} one of the adopted EoSs describing
NSs was the SLy EoS, which is still
consistent with current observations~\cite{Abbott:2018exr,Miller:2019cac}. Here, our starting point is a modification
to this EoS, which we describe with three different piecewise
polytropes, as specified in Table~\ref{tab:eos} and displayed in Fig.~\ref{fig:eoses}. We make this modification
so that the addition of a PT to the EoS requires only four polytropes, which is the default choice of the code, and refer to this fiducial EoS without the addition of a PT as ``SLy.''
Note that with these four polytropes it is already possible to parameterize
a wide variety of first order phase transitions in the EoS, as shown in Fig.~\ref{fig:eoses}.

\begin{table}[h]
   \centering
   \caption{Characterization of the different EoSs used in this work. Each EoS is defined as a piecewise polytrope with $n=4$ segments and with $ K_0[CGS] =
      3.59389\times 10^{13}$ and $\Gamma_0 = 1.35692$.
      Each segment is delineated by a transition density $\rho_i$
      expressed in \texttt{cgs} units.
      Note that for the modified SLy, polytropes 1 and 2 have the same value of
      $\Gamma_i$ and only the lettered EoSs have a segment with $\Gamma_i=0$.
   }
   \begin{tabular}{ccccclll}
      \hline
      EoS & $\Gamma_1$ & $\Gamma_2$ & $\Gamma_3$ & $\log_{10} \rho_0$ & $\log_{10} \rho_1$
      & $\log_{10}\rho_2$
      & $\Delta_\rho(\times 10^{14})$ \\
      \hline
      SLy  & 2.9965  & 2.9965  & 2.851    & 14.165  & 14.7     & 14.9367 & ---\\
      A    & 2.9965  & 0.0     & 2.851    & 14.165  & 14.9367 & 15.1     & 3.95\\
      B    & 2.9965  & 0.0     & 2.851    & 14.165  & 14.9367 & 15.3     & 11.3\\
      C    & 2.9965  & 0.0     & 2.851    & 14.165  & 14.95    & 14.97    & 0.42\\
      \hline
   \end{tabular}
   \label{tab:eos}
\end{table}

A piecewise polytrope is chosen such that the pressure is given in terms of
the density by
\begin{equation}
p_\mathrm{cold}(\rho) = K_i \rho^{\Gamma_i}
\end{equation}
where $i$ denotes the particular section of the piecewise function and runs from 0 to 3.
Enforcing continuity determines the constants $K_i$.

The PT consists of a section in which $\Gamma_2=0$ beginning at some onset density $\rho_1$ and extending to some density $\rho_1+\Delta_\rho$, similar to that adopted in Refs.~\cite{Lindblom:1998dp,Chen:2019rja}. For this study, the stiffness of the EoS at densities above the transition ($\rho>\rho_1+\Delta_\rho$) remains the same as that for the SLy EoS above $\rho_1$, although in principle we are able to modify this as well. 

We consider here just a few such variations. The parameters for each EoS are
listed in Table~\ref{tab:eos}, and we display them in Fig.~\ref{fig:eoses}.
In particular, we show the pressure as a function of density which
defines a barotropic EoS. We also show in the right panel the family of
isolated, spherically symmetric stellar solutions which the EoS generates
from solving the TOV equations with \texttt{Magstar} from the \texttt{Lorene}
package~\cite{lorene}. 
One generally expects a change in stability at the extrema of such 
mass-radius curves. Thus, stars near the region at which the PT occurs
(near the cyan circle) are expected to be stable for EoSs A and C, but not B,
and the evolutions discussed below (Section~\ref{sec:isolated})
are consistent with this expectation.
Also shown is  a cyan circle indicating the particular star the evolution of which is discussed below in Section~\ref{sec:isolated}.

Besides the piecewise polytrope, an additional {\em thermal} component of the EoS is included
during the evolution to account for the thermal component of the fluid, 
represented as an ideal fluid with $\Gamma_\mathrm{thermal}=1.75$. Therefore, the pressure and the internal energy $\epsilon$
have two components
\begin{eqnarray}
\label{piecewise_p_eps_cold}
p &=& p_\mathrm{cold} (\rho) + (\Gamma_\mathrm{thermal} -1) \,\rho\, \epsilon_\mathrm{thermal}  ~~,~~ \\
\epsilon &=& \epsilon_\mathrm{cold}(\rho)  + \epsilon_\mathrm{thermal}.
\end{eqnarray}
At the initial time, $\epsilon_\mathrm{thermal}$ vanishes
because the initial data relies only upon the piecewise polytrope with $p = p_\mathrm{cold}$.
We monitor this thermal component in representative cases as another indication
of the dynamics.

Along with the pressure and the thermal component, the speed
of sound as a function of density is calculated for the initial data as 
\begin{equation}
c_s^2 = \frac{\Gamma_i \, p}{h} 
\label{eq:csq}
\end{equation}
where $h = \rho \left( 1+\epsilon\right) + p$ is the enthalpy. The sound speed is shown
in Fig.~\ref{fig:eoses} in the same frame as the pressure. The effect
of the phase transition is to decrease the maximum speed of sound attained.
Dotted black lines in the left panel of Fig.~\ref{fig:eoses}
indicate the region within which all the EoSs here are causal ($c^2_{s}\le 1$),
and we note that our evolutions do not probe above this high density regime.

Despite the phase transition causing the EoS not to be a 1-to-1
function of density, our inversion from conservative
to primitive variables causes no numerical problems. We note
that the total pressure has both cold and thermal components, and
is still a uniquely defined function of the density and internal energy.
In particular in solving the transcendental function needed for
the inversion, only $p_\mathrm{cold}$, not its inverse  appears,
which is unambiguously well defined.

For the magnetized isolated stars studied in this paper, we assume an initially poloidal magnetic field confined to the stellar interior and calculated from the  vector potential 
$A_ {\phi} \propto r^2 (P - P_\mathrm{cut})$, where $P_\mathrm{cut}$ is a hundred times the pressure of the atmosphere  (about $2\times 10^{32}$ dyn/cm$^2$) and $r$ is the distance to the rotation axis.  The maximum magnetic intensity at the center is $6\times 10^{13}$ G. We then
have the freedom to rotate this configuration some angle, $\theta$, with respect
to the rotational axis (here we choose $\theta=10^o$).

\section{Results}
We discuss first the evolutions of isolated stars (spinning or not). Although observable
effects are much more likely to arise from the dynamics of binaries than
isolated stars, it is informative to study the dynamics of stars in
isolation as a first step to fully understand the problem. We have performed evolutions with 
different resolutions which indicate that the results presented are consistent and within the convergent regime.
For example, the finest level covered the isolated stars with approximately $77$, $116$, and $155$ points across the star. 
Mass was conserved with high accuracy, varying less than
a percent during our simulations and improving with resolution.
We follow this with a discussion of binary mergers.

\subsection{Isolated NSs}
\label{sec:isolated}

Past studies of the evolution of individual stars generally observe
stable stars to oscillate around their static solutions because of slight 
numerical perturbations, or, if unstable,
to collapse to black holes. Indeed these two outcomes are observed here.
However, to force more significant dynamics, we increase the artificial
atmosphere (introduced as part of the standard method of solving
for relativistic hydrodynamics) by six orders of magnitude above the
usually small level chosen (roughly at the level of $0.3 \rho_\mathrm{nuclear}$).
We stress we do not consider such a scenario physically generic,
but, in particular, this choice allows us to 
explore the extent to which possible accretion from a companion
can trigger interesting dynamics in a star with an EoS allowing for a PT.

This atmosphere induces strong accretion by the NS (estimated at $\dot M \simeq 10^{-8}$).
Several important features
are observed as a consequence: (i)~strong density waves are produced that, as they reach
the stellar surface, both reflect an inward propagating wave and expel a thin layer of fluid,
(ii)~this outer layer sweeps against the atmosphere, essentially halting further
accretion for a significant fraction of the stellar dynamical time, and
(iii)~the propagation speed of such waves is changed significantly 
in the region where the PT
takes place, producing other waves at the interface. 
The combined
impact of these effects plays a strong role regulating the longer term behavior of the
star as we discuss below.

\begin{figure}
	\includegraphics[width=3.5in]{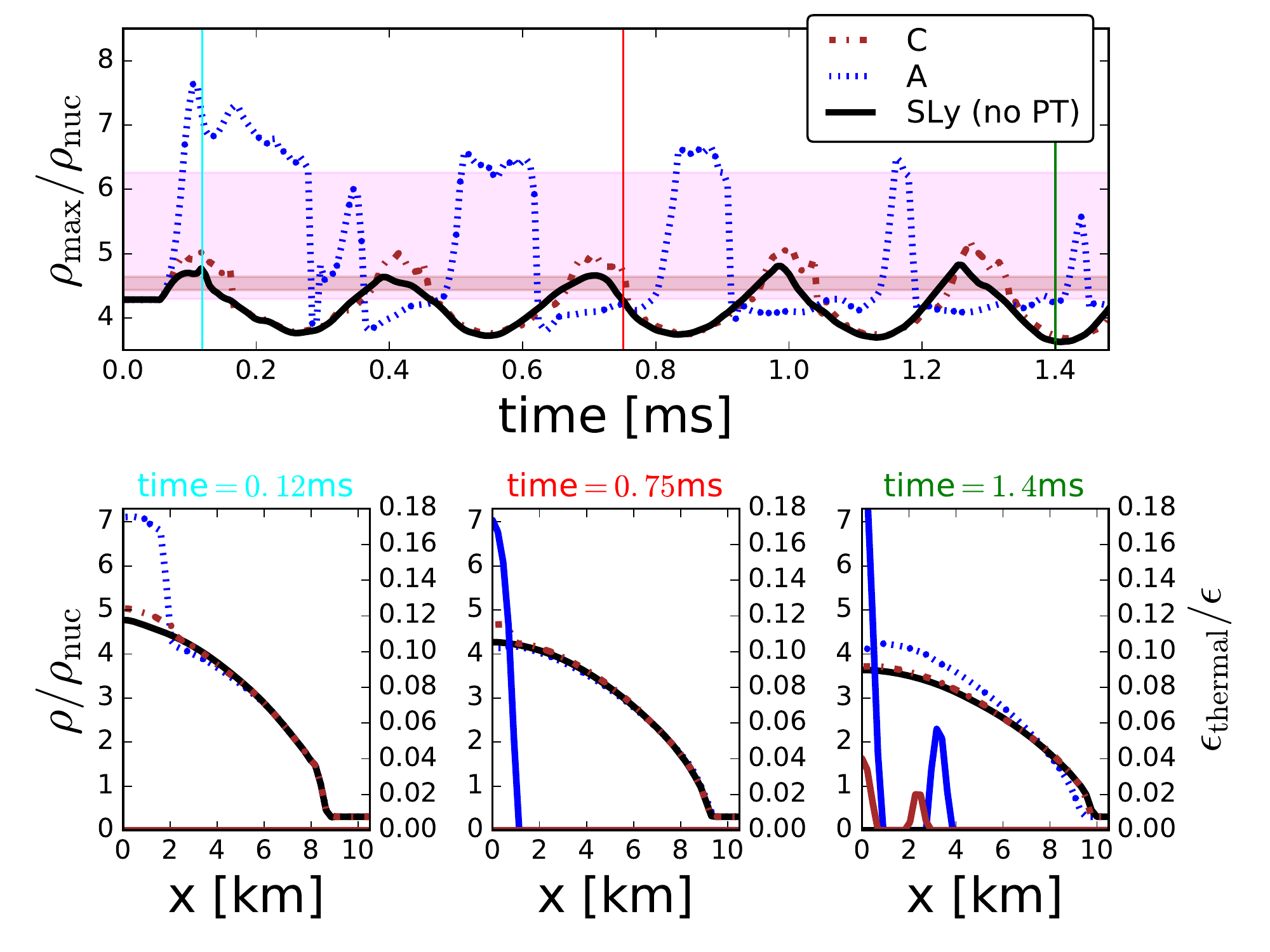}
	\caption{Dynamics of isolated, non-rotating stars.
   The star represented by a cyan circle in Fig.~\ref{fig:eoses} is
   evolved with three different EoSs, and the maximum of the density
   as a function of time is shown (\textbf{top}). 
   Also shown are two horizontal bands indicating the regions of
   the PT ($\rho_1 \le \rho \le \rho_1+\Delta \rho$) for EoS A (magenta)
   and EoS~C (brown).
   For the SLy EoS with no PT (solid black), the star oscillates as expected.
   For the EoS A (dotted blue), as the density increases, it shoots past the
   amplitude of the star with no PT, indicative of the core changing phase.
   These excursions to hybrid phase ultimately settle back to the hadronic star.
   At \textbf{bottom}, profiles of the density along the positive $x$-axis are shown
   for the three times indicated by vertical lines in the top frame. In the first
   of these, stars described by EoS A and C have non-hadronic cores.
   Also shown with the scale on the right is $\epsilon_\mathrm{thermal}$ representing the fractional
   thermal component of the internal energy. These curves can be differentiated
   from the density both because they are all drawn with solid line and because the thermal energy differs from zero only well inside
   the star.
   At early times, the internal energy
   is entirely comprised of the contribution from the cold component.
	}
	\label{fig:singlestar}
\end{figure}

In Fig.~\ref{fig:singlestar}, we characterize the dynamics of one
particular non-rotating star for three different EoSs. The initial data
for this star is
indicated with a cyan circle on the right
panel of Fig.~\ref{fig:eoses}, and, because its maximum density is
below the PT onset density ($\rho_1$ for EoS A and C), it is a solution
for all three EoSs.

The top frame shows the maximum density as a function of time for the
three evolutions. With no PT, the black curve shows the density oscillate
as the star is `pushed' by the large atmosphere. In contrast, the two
evolutions with a PT display large excursions in density consistent with
undergoing a phase transition in the core. For each of the EoSs with
a PT, we show the density range over which the PT occurs by a horizontal
band, magenta for EoS A and brown for EoS~C. These excursions in maximum
density generally send the maximum density to a value slightly above
each density range.

It is important to realize that only some inner core undergoes the PT,
and we show in the bottom frames of Fig.~\ref{fig:singlestar} the radial 
profile of the density at three different times. The earliest such profile,
when the densities are at their peaks, shows clearly the region containing non-hadronic matter, roughly a region
out to 2 km. In some of the later excursions, instead of a convex core
of non-hadronic fluid, a small, irregular shell forms such that the
density maximum is no longer at the center of the star.

The first panel on the bottom of Fig.~\ref{fig:singlestar}
might be misinterpreted as indicating that these stars have 
different masses despite beginning with the same initial data. A  closer look at the total mass, calculated far away from the source, indicates that this is not the case. Furthermore, a simple model can
explain how these profiles characterize stars with the same mass. Comparing the profiles for EoS~A and SLy, we assume:
(i)~the non-hadronic core extends to a radius of roughly 2\,km while both stars have radius 8\,km, (ii)~the non-hadronic core has twice the density of the SLy star, and, finally, 
(iii)~the SLy star has some constant density everywhere, while the EoS~A star
has a different constant density except in its core.
It is straightforward to check that, setting their masses equal and under these assumptions, an overall decrease of roughly 2\% in density (if over the entire star) would allow the star to double the density of its inner core, and
this difference is roughly consistent with what appears in the figure.

It is interesting to contrast the dynamics of the SLy star with that of EoS A and
to wonder why the PT excursions appear damped.
As already noted, the generation of thermal energy with EoS A does not appear
to play a significant dynamical role given the similarities of the two solutions 
in the middle frame of Fig.~\ref{fig:singlestar} despite the fact that one
has thermal energy present.

One can also consider numerical dissipation, unavoidable in such evolutions.
Running at increasing resolutions decreases the expected dissipation, and 
such tests indicate that dissipation is small and develops slowly in time. Indeed, 
the size of the non-hadronic core shrinks quickly with each excursion even in the higher 
resolution runs. Hence, it cannot explain the damped excursions at early times.

If one looks at the radius of the stars as a function of time, one sees that the 
average radius for both stars increases by about $5\%$ in the first millisecond.
Looking at the right panel of Fig.~\ref{fig:eoses}, this increase indicates that the star
now oscillates around a slightly different star (slightly smaller gravitational
mass) than the one providing the initial data.
As the SLy star expands and 
contracts, its kinetic energy oscillates as well.
The
EoS A star, on the other hand, has both much less kinetic energy and a less
coherent oscillation of it. Instead, the energy provided by the expansion of
the star presumably ends up in the non-hadronic core. As the core gets smaller,
so does the amount of kinetic energy liberated by exiting the PT. The
dearth of kinetic energy at late times in the EoS A star can be seen in
the late-time periods in which the maximum density is relatively constant.

At these late times, the EoS A star lacks the coherent motion and kinetic
energy present in the SLy star, presumably due to the increased occurrences
of characteristics crossing 
(regions which are handled with a high-resolution shock capturing scheme)
as a result of the continuing PTs and the resulting disparate propagation speeds.

We note one final aspect about these non-rotating stars. Our discussion
begins with a comment about use of a large atmosphere to force the dynamics,
and so here we comment on what is instead seen with the normal, small atmosphere
for these stars. The stars once again oscillate, but the amplitude of
the oscillations of radius and maximum density instead have amplitude roughly a fraction
of a percent. The dynamics for SLy, EoS A, and C are essentially the same with
one exception. The maximum density of the star described by EoS A does demonstrate
some excursions into the range $\rho_1  < \rho < \rho_1 + \Delta \rho$ but fails
to exceed this range. These excursions only occur for some intermediate times once
the amplitude of the oscillations have developed and before the star has settled.

We also consider rotating and magnetized stellar evolutions, by setting a small poloidal magnetic field, slightly misaligned with respect to the rotation axis (i.e., $10$ degrees), only in the interior of the star. These evolutions
also incorporate the same large atmosphere and grid structure as the non-rotating
solutions just discussed.

\begin{figure}
	\includegraphics[width=3.5in]{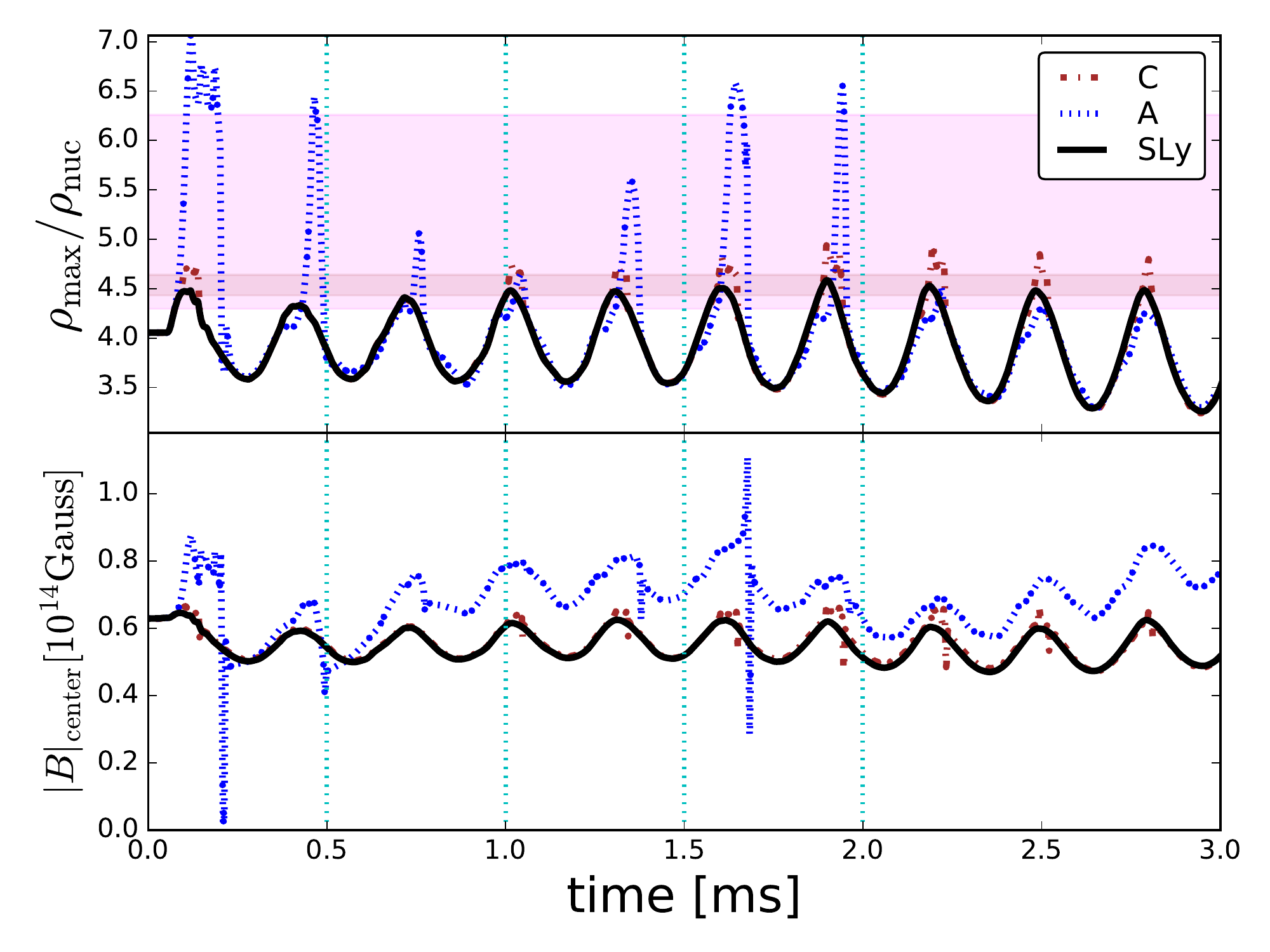}\\
	\includegraphics[width=3.5in]{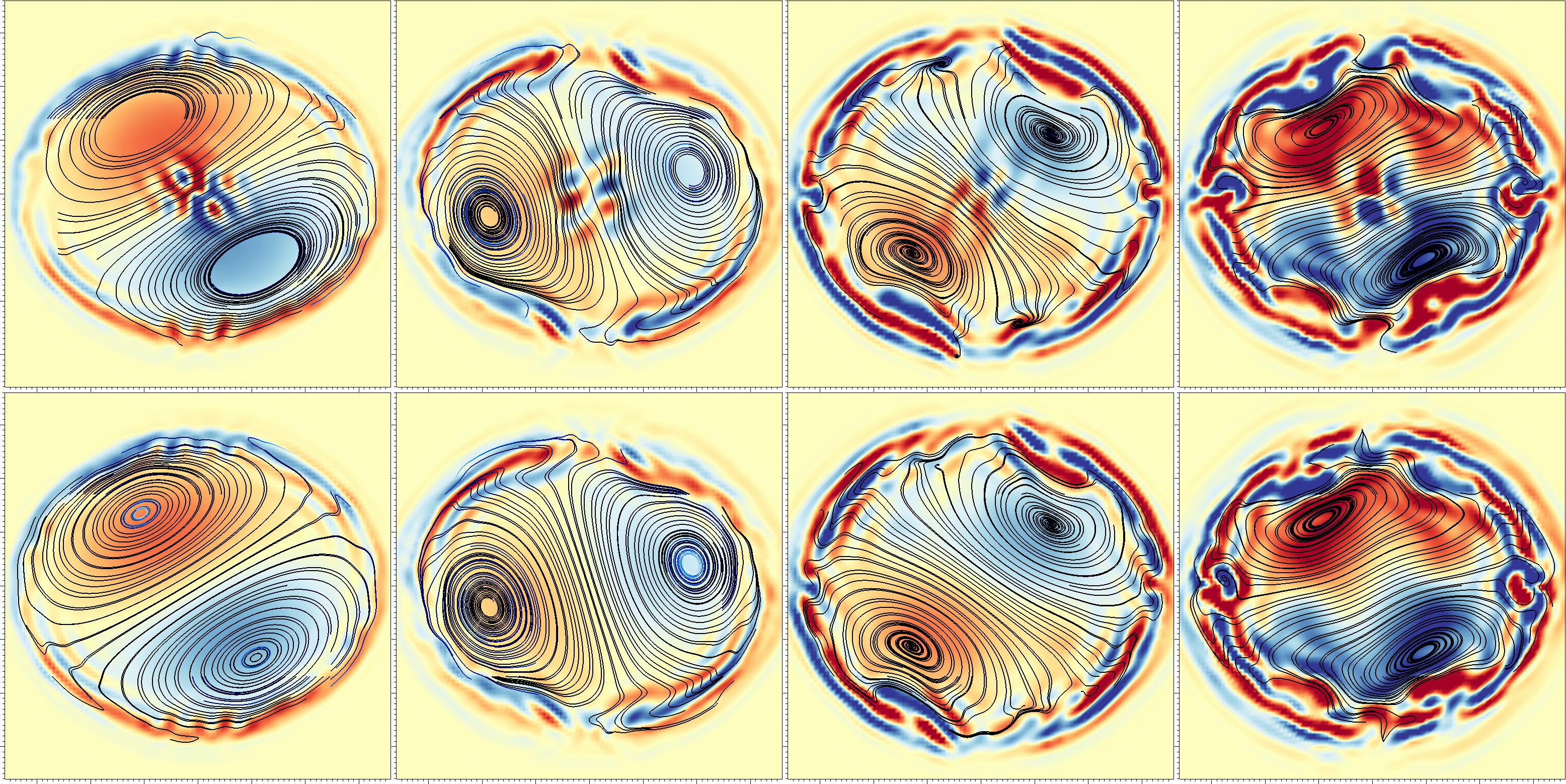}
	\caption{Dynamics of a particular isolated, rotating, magnetized star.
   This star has an initial central density a bit lower than the lowest
   density star shown in Fig.~\ref{fig:singlestar} and rotates at $800\,\mathrm{Hz}$.
    \textbf{Top:} Maximum density as a function of time. The star with
    no PT (black solid) shows the usual stable oscillation.
    The solid, horizontal, magenta line indicates the onset density of the PT.
    \textbf{Middle:} The magnitude of the magnetic field at the center
    as a function of time.
    \textbf{Bottom:} Snapshots of the magnetic field on a meridional plane at times 0.5, 1.0, 1.5, and
   2.0 milliseconds (from left to right). The top row of these panels shows EoS~A and the bottom row shows the SLy star. The colors indicate the curl of the magnetic field all with the same colormap, and the black lines indicate the magnetic field.
    Dotted cyan
    vertical lines are shown in the upper panels indicating the times at which
    these snapshots are taken.
   Although the maximum of the magnetic field between the two EoSs looks similar,
   the structure in the core is significantly different. The passage of the EoS A star 
   through the phase transition appears to contract the core and result in 
   changes to the magnetic field in the core region. However, these changes 
   do not appear to propagate to the surface.
	}
	\label{fig:singlestarrotatingwB}
\end{figure}

As we discuss later, the onset of the PT introduces non-uniform rotation in the interior of the star which
distorts the field topology.  However, this distortion takes place deep inside the NS and the matter pressure
keeps it confined. As a result, there is essentially no impact on the field behavior within the outer envelope of the NS which
implies no direct observational consequences.
In Fig.~\ref{fig:singlestarrotatingwB}, we once again show the maximum density, and we also show the central magnitude 
of the magnetic field for a star rotating
at a frequency of $800\,\mathrm{Hz}$ and with initial central density a bit
lower than the onset density. 
The plot demonstrates a significant difference in magnetic field strength at the center of the star. However, the snapshots of the magnetic field on
a meridional plane at four different times shows that the changes within the core do not propagate to the surface.
We compare the base EoS with no PT with those of EoS A and C which have a PT but are otherwise stable. When looking at the density, one sees behavior
similar to that for the non-rotating case shown in Fig.~\ref{fig:singlestar}. That is, the density for each star with the PT has large vertical excursions as the stellar core undergoes the PT. However, the  oscillations return the star periodically to a purely hadronic star.

The apparently dramatic excursions of maximum density at late
times actually represent
the small scale dynamics of the core in which a small region reaches the onset
density of the EoS.
The cessation of the excursions around $t\approx 2$ ms, however, appears
at roughly the time at which the maximum density drifts downward away from
the onset density. This drift occurs as well for the star with no PT
suggesting that the large scale dynamics is for the star to move towards a different
equilibrium. 

The bottom panel of Fig.~\ref{fig:singlestarrotatingwB} shows significant differences in the central magnetic field  
magnitude.
Although the magnetic field is likely not affecting the dynamics, it is
responding to the changing stellar structure resulting from the transformation
of the stellar core to a non-hadronic state. However, we also show the magnetic field lines along a meridional plane
at a few different times.
The changes in the magnetic field configuration apparent in the figure occur only in the inner part of the star and 
remain confined to that region as pressure overwhelms any potential propagating effect.
On time scales longer than the milliseconds of these evolutions, 
say of order seconds, such differences might potentially reach 
the surface. However, the magnitudes of the differences, not
so large even in these evolutions, would likely
be diminished further as they reach the surface.
Without changes to the surface magnetic field, the ability to observe a NS experiencing a PT
electromagnetically is likely very limited.

\subsection{Binary NSs}

\begin{table*}[t]\centering
	\begin{ruledtabular}
		\begin{tabular}{lllllllllllll}
			EoS & $q$ 
			& $M_{0}^{\rm ADM} $ 			
			&  $m_{b}^{(1)}, m_{g}^{(1)}$ &  $m_{b}^{(2)}, m_{g}^{(2)}$  
			& $R^{(1)}$  & $R^{(2)}$ 
			& $C^{(1)}$  & $C^{(2)}$  
			& $k_2^{(1)}$ & $k_2^{(2)}$  
			& $\kappa_2^T$ 
			& $f_{2}$ \\ 
			
			&  & ~[$M_{\odot}$] &  ~~[$M_{\odot}$]  
			& ~~[$M_{\odot}$] 
			& [km] & [km] &  
			& 
			&  &  & & [khz]   \\ 
						
			\hline
			SLy  & 1.0  &  2.471  & 1.37, 1.20 & 1.37, 1.20 & 11.46 & 11.46 & 0.1607 & 0.1607 & 0.1028 & 0.1028 & 119.9
			& 3.02
			\\ 
			C  &  &   &  &  &  &  & &  &  &  & 
& 3.27
\\
			SLy  & 0.92  & 2.373 & 1.37, 1.20 & 1.25, 1.10 & 11.46 & 11.49 & 0.1607 & 0.1480 & 0.1028 & 0.1122 & 154.4
			& 3.08
			\\
			C  &  &   &  &  &  &  & &  &  &  & 
& 3.02
\\
			SLy  & 0.86  & 2.637 & 1.60, 1.37  & 1.35, 1.18 & 11.43 & 11.46 & 0.1850 & 0.1586 & 0.08484 & 0.1044 & 81.23
			& 3.22
			\\
			C  &  &   &  &  &  &  & &  &  &  & 
& 3.56
\\
		\end{tabular}
	\end{ruledtabular}
	\caption{Summary of the neutron star binaries studied here.
		The initial data were computed using the {\sc Bin star}
		solver from the {\sc Lorene} package~\cite{lorene}. All the binaries  start from an
		initial separation of $37.7$~km. The outer boundary is located at
		$756$~km and the highest resolution level covers both stars with
      a resolution of  $\Delta x_{\rm min}=100$ m. 
      The table displays the mass ratio of the binary
		$q \equiv M_1/M_2$, the baryon (gravitational) mass
		of each star $m_b^{(i)}$ ($m_g^{(i)}$), its circumferential radius $R^{(i)}$ and
		its compactness $C^{(i)}$ (i.e., when the stars are at infinite
		separation), the tidal Love numbers of the individual stars, the polarizability parameter of the binary, and the main GW frequency
		$f_2$ of the post-merger remnant (displayed in Fig.~\ref{fig:binaries_f2_fit}).
    Note that binaries with EOSs~A and~B are not listed because they
    quickly collapse to black hole.
	}
	\label{table:equal_mass}
\end{table*}

We study three particular binaries constructed by \texttt{Lorene} to
be in a quasi-circular orbit. 
These runs use six levels of resolution, the first 5 of which
are fixed with a ratio of two between resolutions. The last level is 
dynamic, tracking the stars, with a refinement ratio of four.
The domain extends in each axis from $\pm 752$ km. 
The finest resolution runs consist of $241^3$ points on the coarse level
(with a finest level grid spacing of $\Delta x=98$ m) and others use
$193^3$ (finest level $\Delta x=123$ m). For just a few runs
and for shorter periods to test that we were in the convergent
regime, runs used up to $343^3$ points.
We note that these binary simulations do not use the large atmosphere
adopted for the isolated stars and do not have any magnetization.
Certain other details about these binaries
are summarized in Table~\ref{table:equal_mass}.
Note that these binaries have different total masses which
complicates direct comparison among the evolutions. 
Future work includes constructing binaries maintaining certain parameters,
but these variations in total mass allow us to scan a broad
range of mass ratio.

We construct an equal mass binary with stellar central densities a bit
below the onset density for the EoSs considered here. Once again we consider evolutions with EoSs A, B, C,  and SLy (without a PT). The initial data, being
below the PT, is the same for all four  runs.
The total gravitational (or ADM) mass of the system is $2.47 M_\odot$ and the  initial orbital angular velocity is $2190$~rad/s.
Each star has a baryonic mass of $M_b=1.37 M_\odot$ and is initially separated from the other by $37.73$~km.

\begin{figure}
	\includegraphics[width=3.0in]{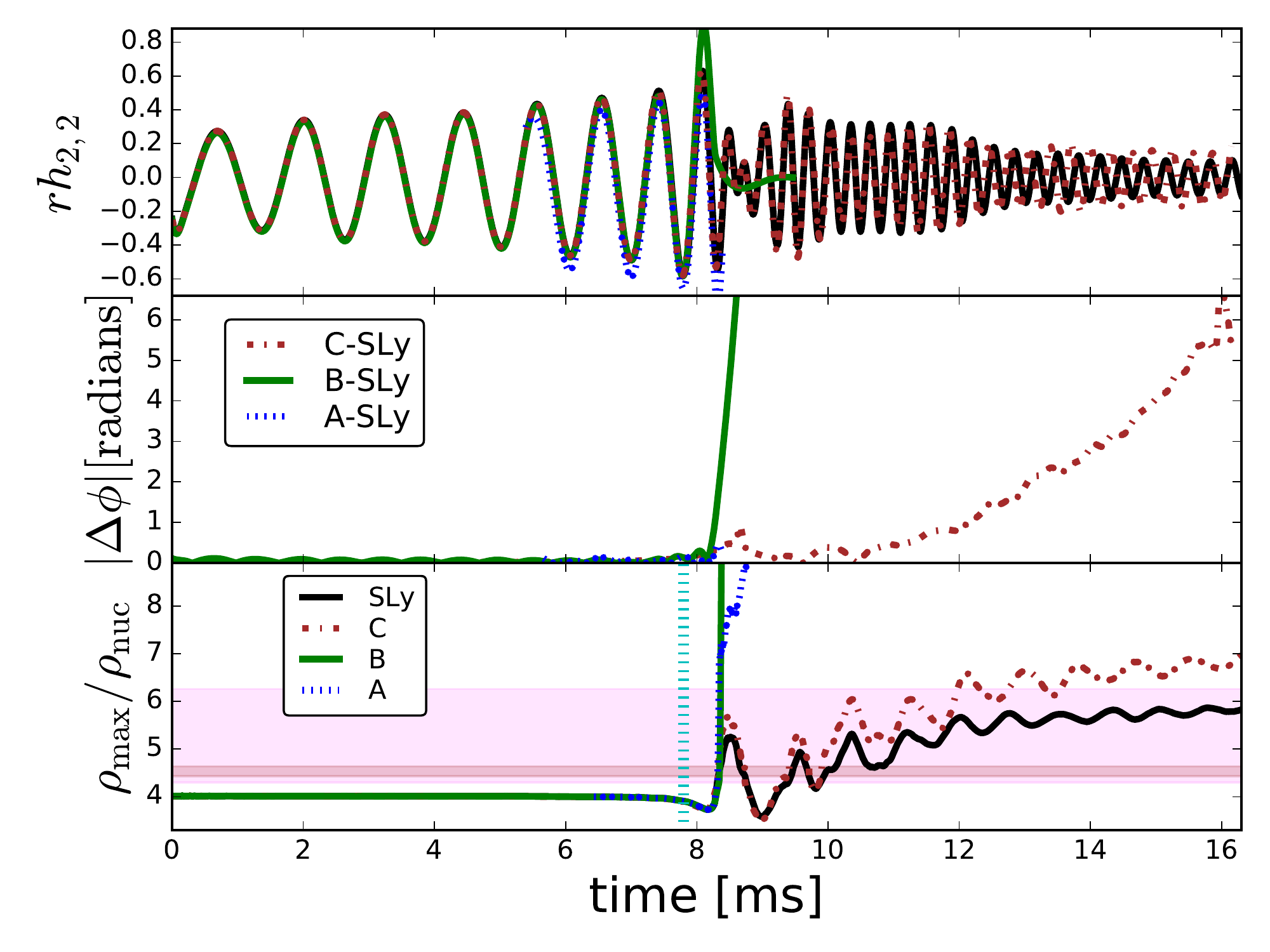}
	\caption{Dynamics of an equal-mass, binary NS merger.
    \textbf{Top:} The real component of the wavestrain as a function of time.
    \textbf{Middle:} The difference in phase with respect to the SLy EoS.
    \textbf{Bottom:} The maximum density as a function of time. The horizontal
       band indicates the density region in which the PT occurs: wide magenta (EoS A) and thin brown (EoS~C). 
   The vertical, cyan line shows the time at which the two stars touch.
    The mergers with EoSs A and C both result in prompt collapse at merger.
    The small phase differences at early times are likely due to the double time
    integration to get the strains and not to any physical effect.
   }
	\label{fig:binary_1.37}
\end{figure}

The evolutions of this binary with different EoSs are displayed in Fig.~\ref{fig:binary_1.37}. The top frame shows the dominant strain mode, $h_{2,2}$ while
the middle frame shows the phase difference for each case compared to the
EoS with no PT. The bottom frame shows the maximum density, $\rho_\mathrm{max}$,
as a function of time. Within this frame are shown the density bands
indicating the PT, thin brown for EoS~C and wide magenta for EoS A.

This
bottom frame indicates that the binaries apparently share the same
evolution at times before merger because the density fails to reach
the density at which the EoSs differ. The reader, however, may notice
small differences in the strain in the top frame and small phase
differences at these pre-merger times. These differences are due to
differences arising from the double time integration of $\psi_4$ over
somewhat different time lengths and boundaries. As such, we consider
phase differences of order 0.1 radians as something of a floor, and expect
physical, and potentially observable, phase differences as those that
exceed this floor value.

The binary with the SLy EoS merges and forms a hyper-massive neutron star which remains stable for at least 10 ms after the collision. 
The cases with a PT depart from this behavior.

It is not at all surprising that the result for EoS~B collapses promptly 
at merger based on the stability properties of the EoS at high mass.
This expectation can be explained by examining the TOV
solutions associated with EoS~B as plotted in the right frame of
Fig.~\ref{fig:eoses}. A local maximum occurs in the mass-versus-radius
plot right near the region describing hybrid solutions and such an
extremum indicates a change in stability; in contrast, the other EoSs do
not have an extremum in that region.

We also observe the binary described by EoS A collapsing quickly
upon merger. 
Again, by  examining the TOV
	solutions in the right frame of
	Fig.~\ref{fig:eoses} one can clearly see that
EoS~A does not support as massive stars as EoS~C, and 
one expects the remnant to be fairly massive.

The remnant under EoS~C survives the merger, and, once the
maximum density increases after the stars merge (see bottom panel of Fig.~\ref{fig:binary_1.37}),
the GW signal shows differences (see the middle panel showing the phase difference
between the GW signals).  However, not until a bit more than two milliseconds after
merger does the phase difference between EoS~C and SLy begin to grow steadily.

\begin{figure}
	\includegraphics[width=3.0in]{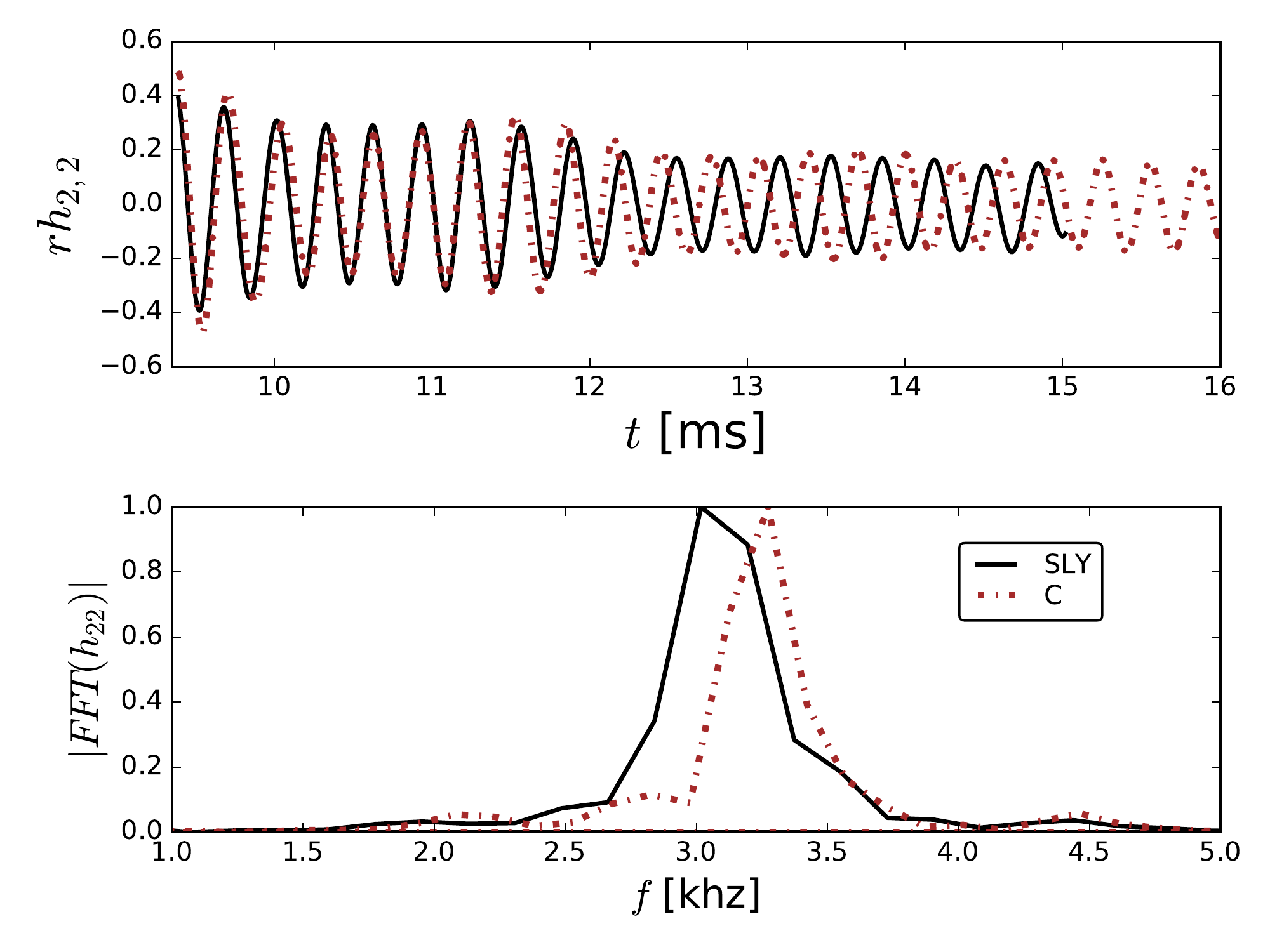}
	\caption{The FFT of the post-merger signals from the equal mass binary mergers shown in Fig.~\ref{fig:binary_1.37}. The power spectral density is shown normalized to the maximum. The remnant that undergoes a PT
   oscillates at a higher frequency than that without the PT, indicating a more
compact remnant.
	}
	\label{fig:binary_fft}
\end{figure}

Because the GW differences occur post-merger, we analyze the frequency
differences with a fast-Fourier-transform~(FFT) of just the post-merger
region of the signal. We show this FFT and the signals in this region
in Fig.~\ref{fig:binary_fft}. The remnant with the PT oscillates at
a higher frequency consistent with the results of Ref.~\cite{Most:2018eaw,Most:2019onn,Weih:2019xvw}. Such a behavior can be understood in terms of a simple model. In the equal mass
case, the PT takes place at a central region in the remnant. There, the density is higher and,
as a consequence of its moment of inertia decreasing (due to approximate mass conservation),
the angular frequency goes up. In cases where such a region is sufficiently large, its contribution
to gravitational waves from the system dominates. Thus, there is a tendency towards higher frequencies due to the PT. 

\begin{figure}
	\includegraphics[width=3.0in]{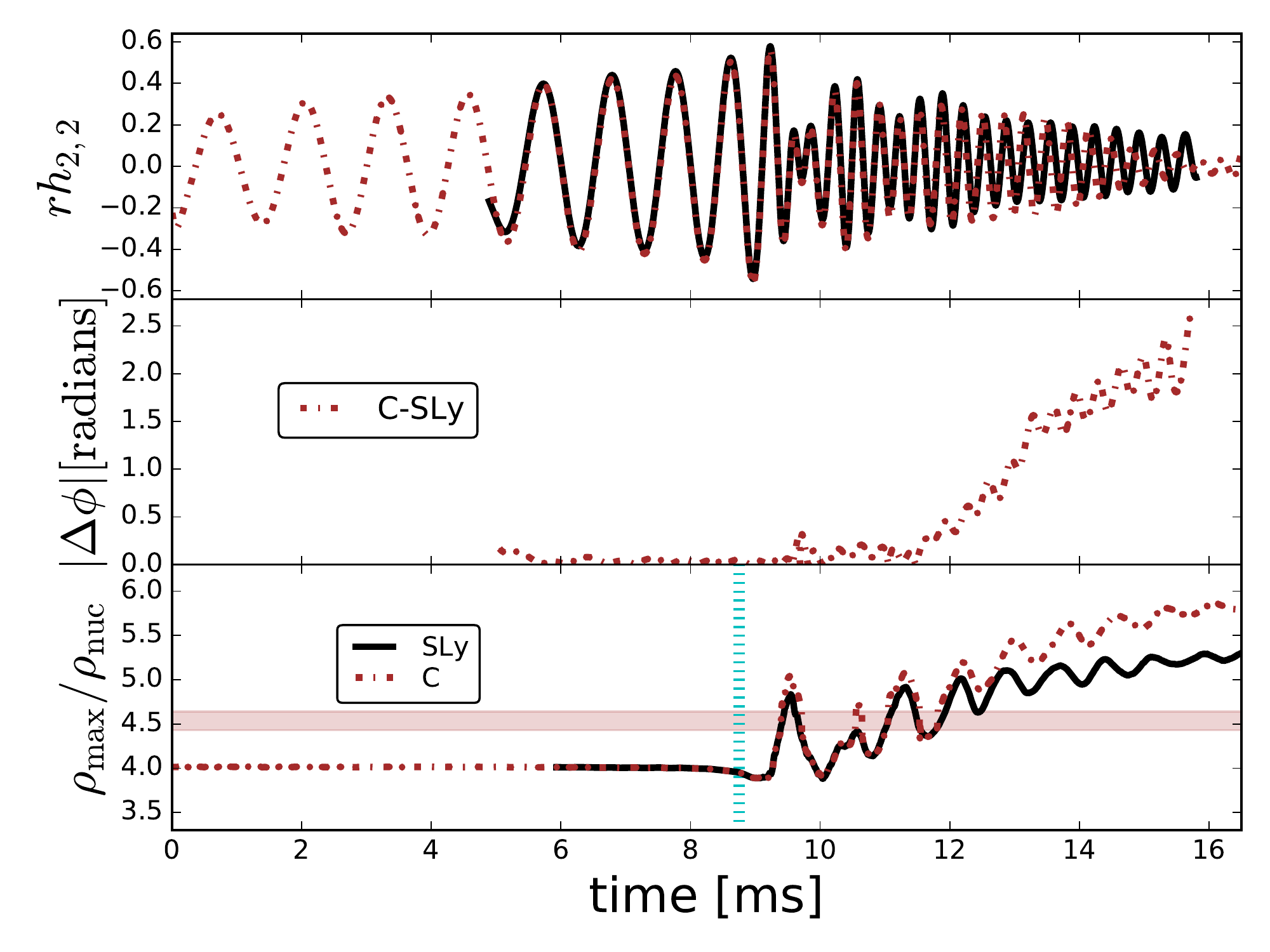}
	\caption{Dynamics of the $q=0.92$ unequal-mass, binary NS merger with masses
     $1.2 M_\odot$ and $1.1 M_\odot$.    
   The vertical, cyan line shows the time at which the two stars touch.
	}
	\label{fig:binary_unequal}
\end{figure}

\begin{figure}
	\includegraphics[width=3.0in]{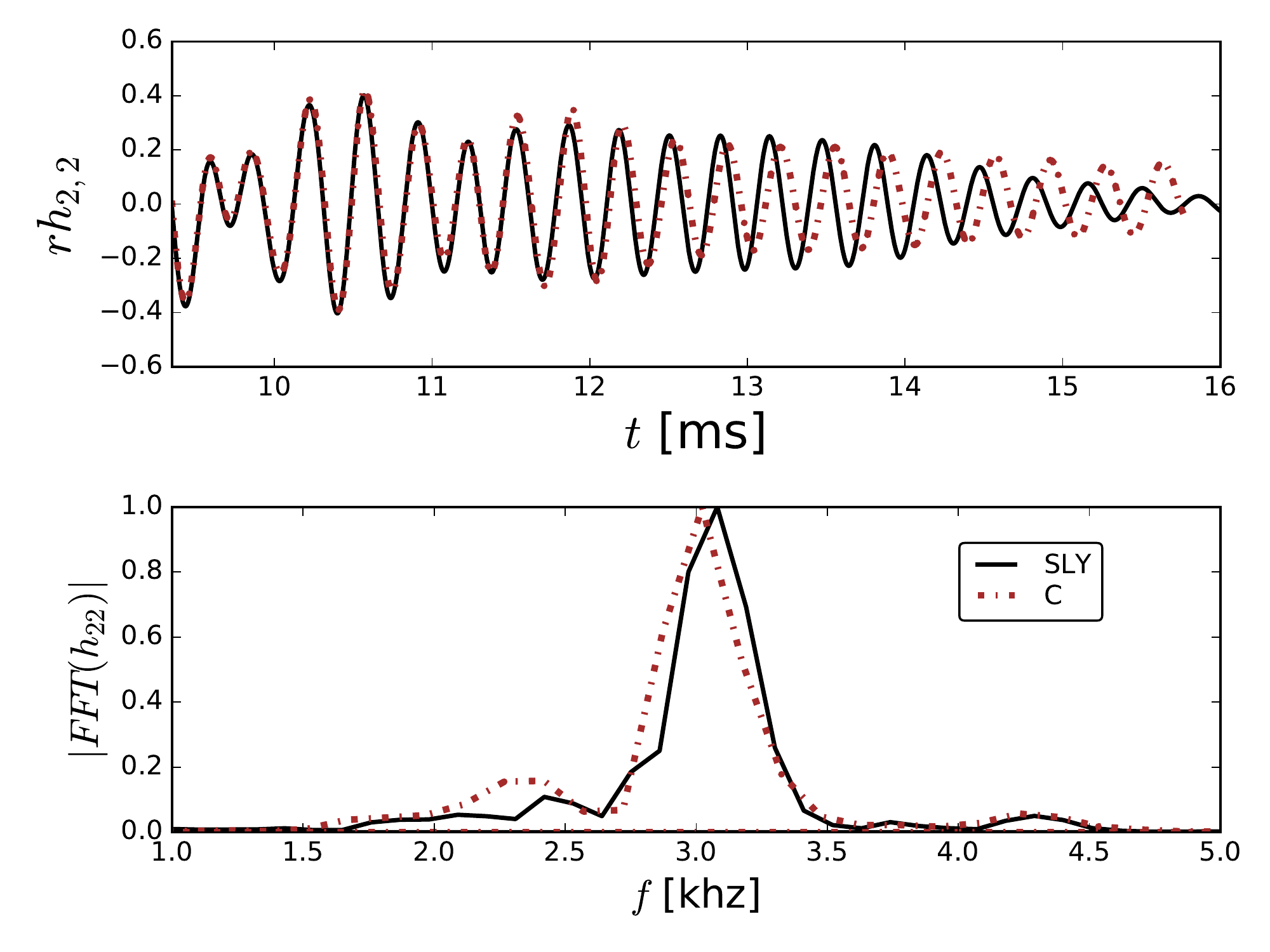}
	\caption{The FFT of the post-merger signals from the $q=0.92$ unequal mass binary mergers shown in Fig.~\ref{fig:binary_unequal}. The power spectral density is shown normalized to the maximum. 
	}
	\label{fig:binary_unequal_fft}
\end{figure}

We next consider unequal mass binaries. The study of asymmetric BNS binaries has become yet more relevant 
and interesting in light of the
recent observation of a BNS with asymmetric mass ratio of $q=0.78$ via pulsar timing~\cite{Ferdman:2020huz}. 
First, we consider one with mass ratio\footnote{We define the mass ratio in
terms of the two gravitational masses of the binary components with
$M^2_g \ge M^1_g$ and $q\equiv M^1_g/M^2_g$.} $q=0.92$
shown in Fig.~\ref{fig:binary_unequal}.

\begin{figure}
	\includegraphics[width=3.0in]{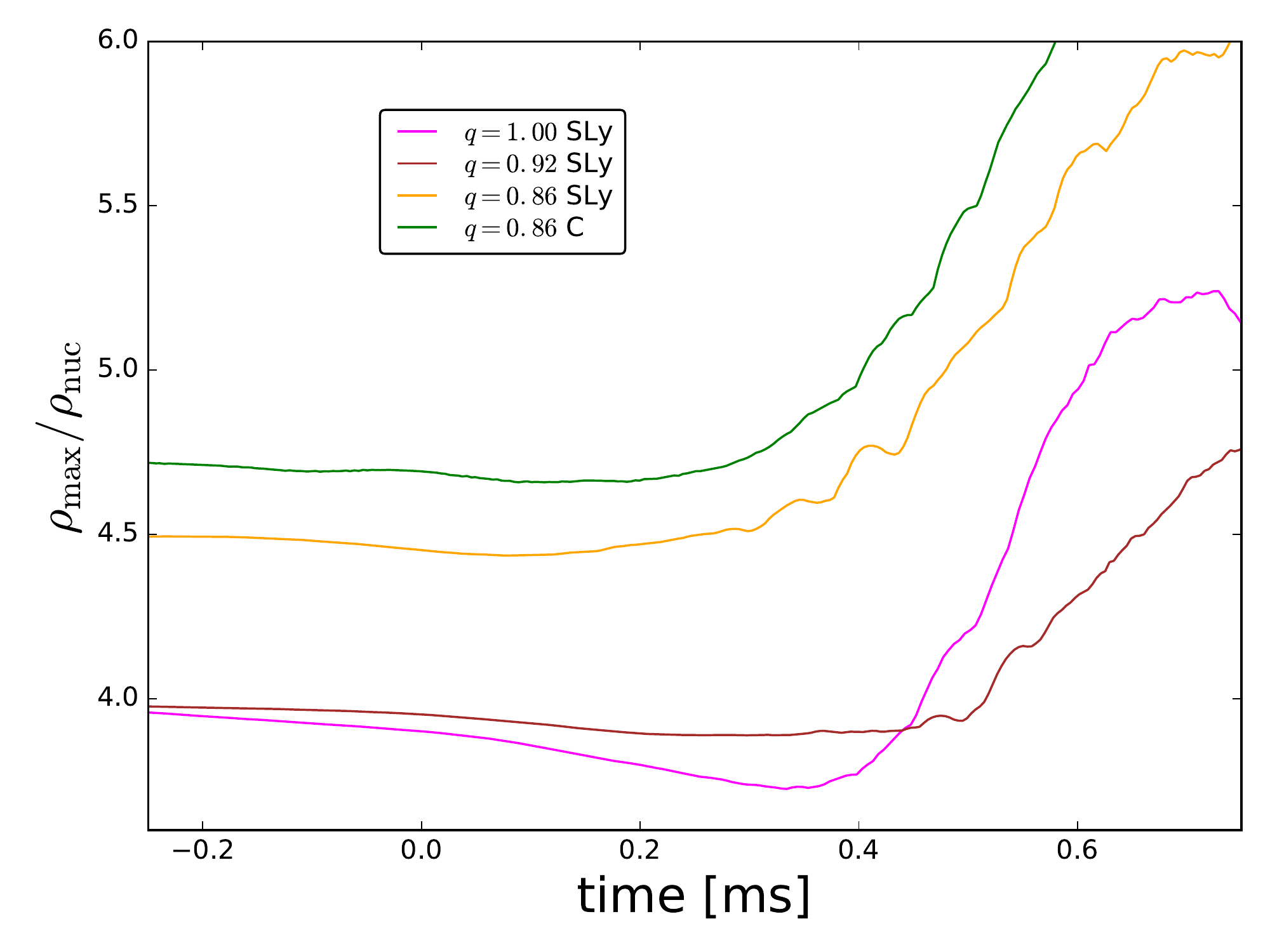}
	\caption{Near-merger behavior of maximum density for the various binaries studied here.
	The times 
   for all binaries have been shifted such that $t=0$ denotes the moment 
   when the stars first touch.
	The maximum density dips soon after the stars touch.
	}
	\label{fig:nearmerger}
\end{figure}

In this case, the overall qualitative behavior is very similar to the equal mass case. Importantly
however, the main frequency in the gravitational waves produced post-merger, as shown in Fig.~\ref{fig:binary_unequal_fft},  is quite close for
both the purely hadronic and PT cases. 
Two factors help explain why the post-merger frequencies are closer together than in the $q=1$ case.
The first is that the non-hadronic core for this remnant is smaller than
that formed in the equal mass case and therefore presumably contributes little to increasing the
rotational frequency via conservation of angular momentum. 
This is related to the fact that this binary reaches the smallest post-merger densities of these binaries (see Fig.~\ref{fig:nearmerger} discussed below).
The other factor is a bit more subtle
and particular to the unequal mass case. Unlike the ``dumbbell'' formed in the early merger of the equal
mass case, the remnant of an unequal mass merger is dominated by the more massive object which happens
to contain the non-hadronic core. As the core gets closer to the rotational axis, its contribution
to GW production decreases and its own quadrupole moment begins to be the dominant effect.

\begin{figure}
	\includegraphics[width=3.0in]{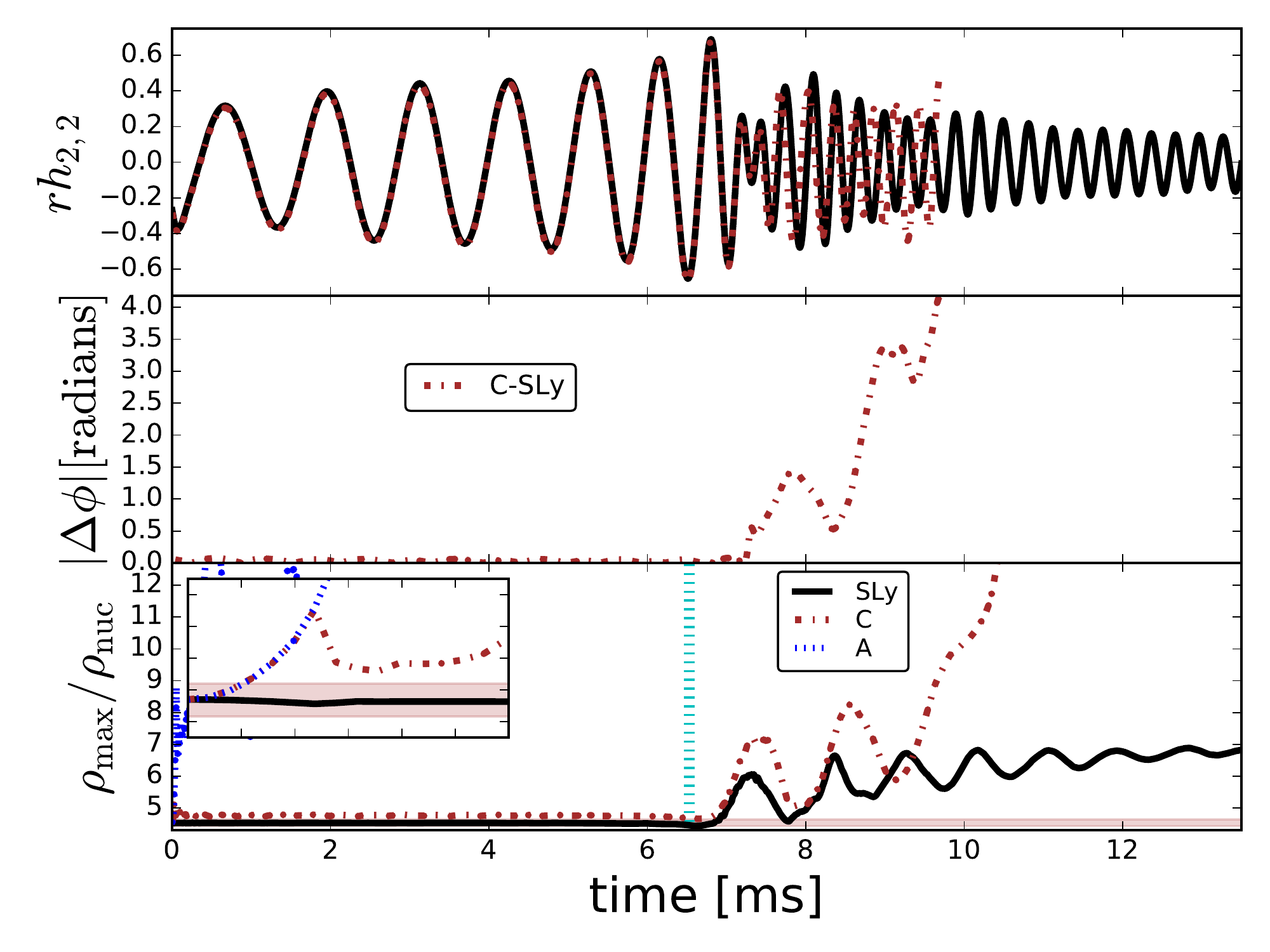}
	\caption{Dynamics of a $q=0.86$, unequal-mass, binary NS merger 
     with masses
     $1.37 M_\odot$ and $1.18 M_\odot$.   
   This binary lacks a PT in the ID because \texttt{Lorene} seems unable
  to produce a binary with one hadronic star and one with a non-hadronic core.
   However, the density of the high mass star is such that the evolution quickly shows it to go through the PT,
   as shown in the inset which covers the period $0\le t \le 0.06$~ms. One is then evolving a binary with one
   hybrid star and one hadronic.
   The vertical, cyan line shows the time at which the two stars touch.
	}
	\label{fig:binary_unequal6ab}
\end{figure}

Next, we study a case with $q=0.86$. The idea
here was to construct a binary with one hybrid star that has a maximum
density above the PT of the EoS while the other is a normal hadronic star
with lower central density. This high mass star could potentially
undergo the PT in the opposite sense as those described above.
However, we were not able to generate such a binary
with \texttt{Lorene}, and so instead we started with a purely hadronic,
unequal mass binary. 
Note, this difficulty arises only for this mass ratio, not the previously presented cases.
However, we stress that the choice of a sufficiently massive star leads to the PT taking place dynamically and the resulting binary is 
of mixed type.
Indeed, once evolved with EoS A or C, the higher mass star quickly undergoes the PT
(see the inset of the bottom panel of Fig.~\ref{fig:binary_unequal6ab} showing
the early time behavior of the maximum density).

\begin{figure}
	\includegraphics[width=3.0in]{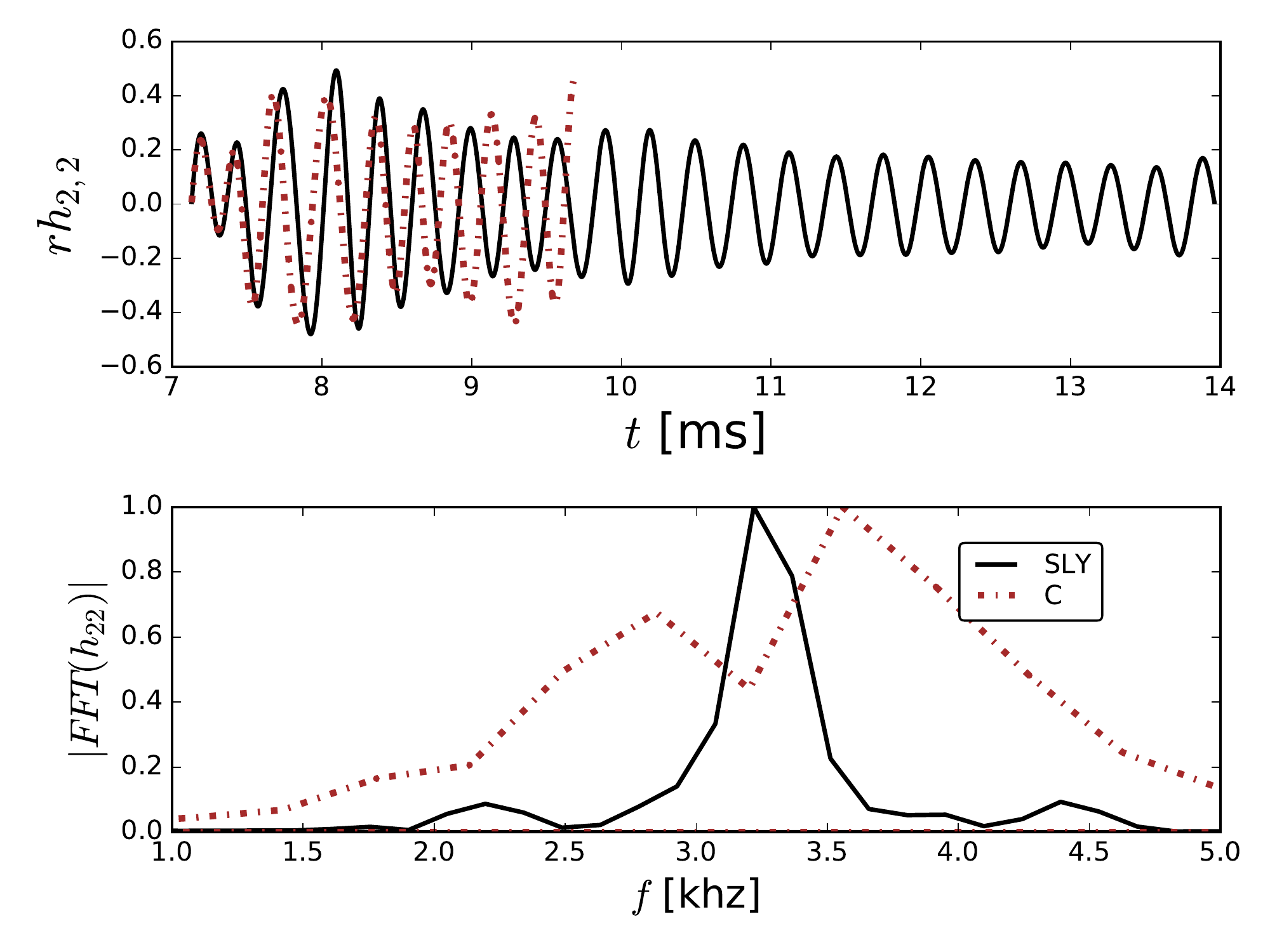}
	\caption{The FFT of the post-merger signals from the $q=0.86$ unequal mass binary mergers shown in Fig.~\ref{fig:binary_unequal6ab}. The power spectral density is shown normalized to the maximum. 
	}
	\label{fig:binary_unequal6ab_fft}
\end{figure}

As shown in Fig.~\ref{fig:binary_unequal6ab}, the maximum density at early
times shows large differences among the three EoSs. With EoS SLy, the binary
evolves without any significant change to the maximum density. In contrast,
both EoS A and C show a quick rise in the maximum density indicating that
the stars are undergoing their respective PT. The high mass star with EoS A
quickly collapses. 
As shown in Fig.~\ref{fig:binary_unequal6ab_fft}, the post-merger differences
resemble those in the equal mass case with the EOS~C remnant oscillating at
a higher frequency.

Common to all three binaries studied here, the maximum density decreases just after the stars touch and
just before merger (see the bottom panels just after the vertical cyan line in Figs.~\ref{fig:binary_1.37}, \ref{fig:binary_unequal}, and \ref{fig:binary_unequal6ab}). This decrease in maximum density for all the binaries considered
here is also shown in Fig.~\ref{fig:nearmerger}.
Such a decrease is already expected from post Newtonian calculations (e.g.~\cite{PhysRevLett.76.4878}) and has
potentially important consequences.
For stars above the PT, the drop in maximum density does not appear significant enough for the star to drop through the PT
to a purely hadronic star {\em before} the stars come into contact (behavior already indicated by PN arguments where
$\delta \rho_c/\rho_c \lesssim 0.3\%$ in e.g. the equal mass case). However, the trend towards a decrease in central density
continues up to $\delta \rho_c/\rho_c \simeq 2\%$ (for the equal mass case, see Fig.~\ref{fig:nearmerger}). Such a decrease could imply a transition back to a purely hadronic
case for stars barely above the PT.  After such transitory density minima, strong density oscillations could potentially have a correlated behavior ``in and out'' of the PT for some time (akin to
the oscillations discussed before in the case of isolated neutron stars). However, this scenario might only
arise within a narrow set of physical parameters.

\begin{figure}
	\includegraphics[width=3.0in]{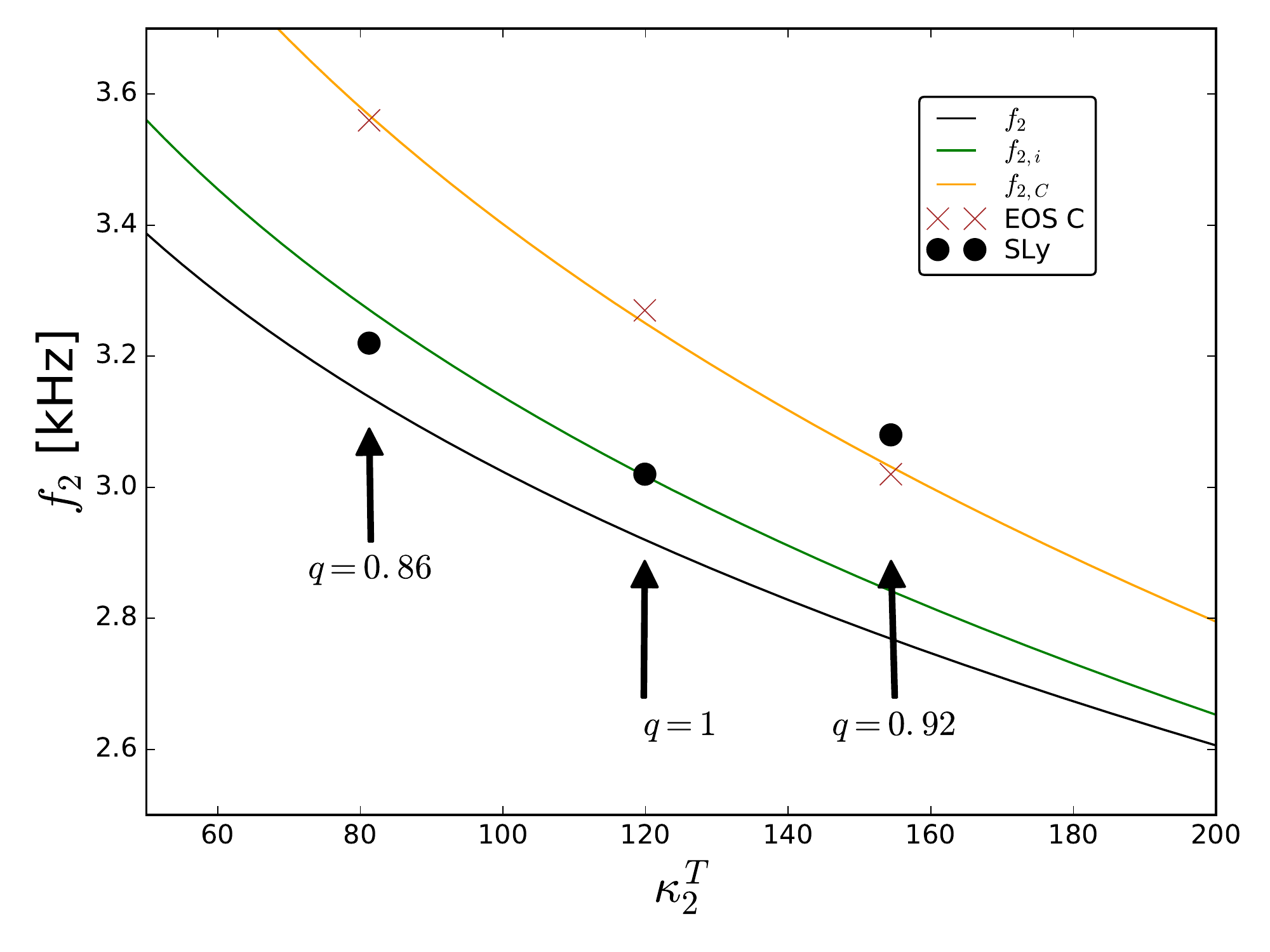}
	\caption{Post-merger $f_2$ frequencies for the various binaries studied here along with fits to these frequencies.
	The fits $f_2    = 5.832-1.118 \left(\kappa_2^T\right)^{1/5}$
	    and  $f_{2,i}= 6.401-1.299 \left(\kappa_2^T\right)^{1/5}$
	come from Ref.~\cite{2016PhRvD..93l4051R}. The frequencies for EoS~C appear to follow
   a similar trend, and a fit to just these three points with the same form
   results in
	         $f_{2,C}= 7.482-1.624 \left(\kappa_2^T\right)^{1/5}$.
	}
	\label{fig:binaries_f2_fit}
\end{figure}

Let us now take a closer look at the dependence of the main frequency, 
$f_2$,  of the post-merger GW signal with the effective ``remnant  
tidal polarizability parameter,''\footnote{Also known as one of a family of ``dimensionless tidal parameters'' in the effective one body approach~\cite{2010PhRvD..81h4016D}.} $\kappa_2^T$, defined as
\begin{equation}
	\kappa_2^T = 2 \left[
	q \left( \frac{X^{(1)}}{C^{(1)}}  \right)^5 k_2^{(1)} + 
	\frac{1}{q} \left( \frac{X^{(2)}}{C^{(2)}}  \right)^5 k_2^{(2)}	
	\right]
\end{equation}	
where $q=M^{(2)}/M^{(1)} \le 1$,  $X^{(i))}=M^{(i)}/(M^{(1)} + M^{(2)})$ and
$C^{(i)} = M^{(i)}/R^{(i)}$,
being $k_2^{(i)}$ the individual tidal Love numbers of each star.
A rather robust functional dependence has been found that relates these two
quantities (e.g. ~\cite{PhysRevLett.112.201101,Lehner:2016lxy,Vretinaris:2019spn}).
Fig.~\ref{fig:binaries_f2_fit} displays
a particular fit of this frequency as a function of this polarizability parameter, obtained by extracting this value from the remnant of many binary neutron star simulations, with different EoS without a PT~\cite{2016PhRvD..93l4051R}. We include on the figure our values for our modified SLy and those obtained with EoS~C computed from the FFTs of the postmerger signals. Because we evolve our binaries just a few milliseconds after merger, we also include the frequency fit $f_{2,i}$ which represents the transient frequency just after merger.

The equal mass ($q=1$) and $q=0.86$ cases show significant differences in the post-merger frequencies, which
could potentially identify
the PT (see e.g. Ref.~\cite{Bauswein:2018bma}).
As noted, the $q=0.92$ case shows a small difference; restricting the FFT to times after 13 ms shows that 
the $f_2$ frequency for EoS~C is arguably consistent to that  of the case without the PT. 

It is not completely clear how complicated the dependence of the post-merger frequency may be when PTs
are involved, and the impact of both mass ratio and total mass of the binary.
More thorough coverage of the parameter space
would be needed to construct a model able to predict the expected values
of $f_2$ in general cases. Despite this need for more coverage, the frequencies obtained for EoS~C
do appear to follow a similar functional form as that obtained for EoSs with no PT, and we present that
fit in the figure.

\begin{figure}
	\includegraphics[width=3.0in]{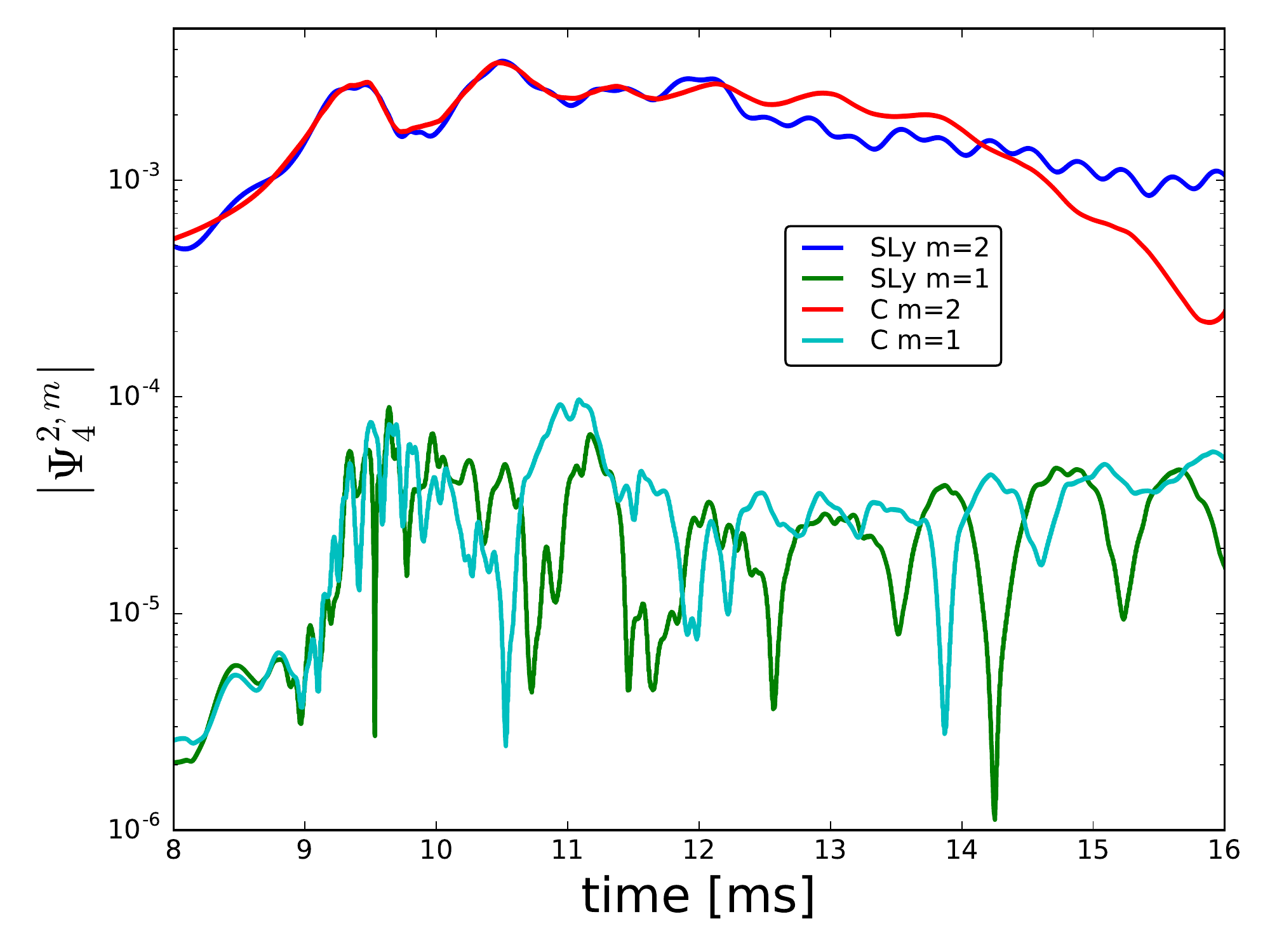}
	\caption{Comparison of the growth of the $m=1$ mode for the $q=0.92$ binary postmerger.
   The growth of this mode, although different between the two EoSs, is 
   roughly comparable. 
	}
	\label{fig:m=1mode}
\end{figure}

Another aspect of the binary merger that may lead to differences due to the presence of a PT is the development
of the $m=1$ mode~\cite{Paschalidis:2015mla,Lehner:2016wjg}.
This mode grows more quickly for
unequal mass binaries, and so the wide range of mass ratios in our simulations does well to probe for
such differences. In Fig.~\ref{fig:m=1mode}, we compare the magnitude of 
the $\Psi_4{}^{2,1}$ mode for SLy and EoS~C with the $q=0.92$ mass ratio,
our longest post-merger evolution and one that typifies the differences in
the other two cases. As shown in the figure, the mode for each case grows qualitatively similarly, suggesting that the PT does not significantly affect the growth rate of this instability.
	
\section{Discussion}

We compare evolutions of NSs and BNSs with EoSs that differ only in
the presence of a somewhat generic PT. That is, the PT is arbitrary and
not motivated by a particular theory of high density matter. Only a
few different PTs are adopted among a large space of possibilities.

With individual NSs, hadronic stars with core densities close
to the PT can undergo the PT and oscillate between hadronic and hybrid
states. The dynamics of this oscillation appears to involve a complex interplay
of a few factors. The star expands and contracts, generates thermal energy,
and all the while the fraction of the core becoming non-hadronic decreases. 
This behavior arises from the accretion and pressure afforded by the artificially high
atmosphere chosen here that is likely unrealistic.
However this interesting behavior might instead be triggered astrophysically.
For example, it would be interesting to assess whether the onset of this behavior
might be induced by strong tidal interactions in an eccentric binary as explored in~\cite{Yang:2018bzx}.

With rotation and magnetic field, the core is similarly dynamic. However,
we observe no significant change to the surface magnetic field.

Evolutions of binary mergers of hadronic stars with densities
close to 
the onset density of a PT, both equal and unequal masses.
show a difference in their GW signatures
at merger when the maximum density reaches the onset density. The
phase difference increases as the post-merger regime continues
and the remnant oscillates at higher frequency than the hadronic
remnant.  We expect that, at least in some cases, these differences may
be observable, although one can imagine possible degeneracies with
EOSs lacking a PT but that otherwise produce condensed cores at high
densities.

We also consider the novel scenario of a BNS composed of a hybrid
star along with a hadronic star. Such a necessarily unequal mass
binary, here with $q=0.86$, is particularly relevant given
the recent detections via pulsar timing of a very unequal ($q=0.78$)
BNS~\cite{Ferdman:2020huz} and with asymmetric compact object binaries by LIGO/Virgo.
A binary such as this offers the possibility that the hybrid star
decompresses and becomes more hadronic dynamically. Although our $q=0.86$ binary
is composed of a hybrid and a hadronic star, this binary collapsed to a black hole
at merger. However, were a PT occurring at smaller density chosen, one could
then construct such a scenario that avoided collapse at merger.

We also computed the dominant post-merger oscillation frequency of the remnant
for these mergers. Future generations of gravitational wave observatories are
expected to have a bandwidth extending to the post-merger regime with the
hope of differentiating a PT in the EoS via these frequency differences. 
Further theoretical analysis of the many outcomes will be required to guide
detection efforts as well as development of efficient ways to search for such subtle 
observables, e.g.~\cite{Yang:2017xlf,Whitaker}.

\bigskip

\begin{acknowledgments}
	
We would like to thank Will East for helpful discussions and to Juan Calderon for addressing our attention to the effects of PT on the $m=1$ instability.
This work was supported by the NSF under grants 
PHY-1912769
and 
PHY-2011383.
CP acknowledges support from the Spanish Ministry of Economy and Competitiveness grants AYA2016-80289-P and PID2019-110301GB-I00 (AEI/FEDER, UE). 
LL was supported in part by NSERC through a Discovery Grant, and CIFAR.  
Computations were performed at XSEDE, Marenostrum and the Niagara supercomputer at the SciNet HPC Consortium. 
Computer resources at MareNostrum and the technical support provided by Barcelona Supercomputing Center were
obtained thanks to time granted through the $17^\mathrm{th}$ (project Tier-0 GEEFBNSM) and $20^\mathrm{th}$(Proposal 2019215177)  PRACE regular calls.  
SciNet is funded by: the Canada Foundation for Innovation; the Government of Ontario; Ontario Research Fund - Research Excellence; and the University of Toronto.
Research at Perimeter Institute is supported by the Government of Canada and by the Province of Ontario
through the Ministry of Research, Innovation and Science.

\end{acknowledgments}

\bibliographystyle{utphys}
\bibliography{paper}

\end{document}

%% file: aas_macros.tex
\def\jnl@style{\it}
\def\aaref@jnl#1{{\jnl@style#1}}

\def\aaref@jnl#1{{\jnl@style#1}}

\def\aj{\aaref@jnl{AJ}}                   
\def\araa{\aaref@jnl{ARA\&A}}             
\def\apj{\aaref@jnl{ApJ}}                 
\def\apjl{\aaref@jnl{ApJ}}                
\def\apjs{\aaref@jnl{ApJS}}               
\def\ao{\aaref@jnl{Appl.~Opt.}}           
\def\apss{\aaref@jnl{Ap\&SS}}             
\def\aap{\aaref@jnl{A\&A}}                
\def\aapr{\aaref@jnl{A\&A~Rev.}}          
\def\aaps{\aaref@jnl{A\&AS}}              
\def\azh{\aaref@jnl{AZh}}                 
\def\baas{\aaref@jnl{BAAS}}               
\def\jrasc{\aaref@jnl{JRASC}}             
\def\memras{\aaref@jnl{MmRAS}}            
\def\mnras{\aaref@jnl{MNRAS}}             
\def\pra{\aaref@jnl{Phys.~Rev.~A}}        
\def\prb{\aaref@jnl{Phys.~Rev.~B}}        
\def\prc{\aaref@jnl{Phys.~Rev.~C}}        
\def\prd{\aaref@jnl{Phys.~Rev.~D}}        
\def\pre{\aaref@jnl{Phys.~Rev.~E}}        
\def\prl{\aaref@jnl{Phys.~Rev.~Lett.}}    
\def\pasp{\aaref@jnl{PASP}}               
\def\pasj{\aaref@jnl{PASJ}}               
\def\qjras{\aaref@jnl{QJRAS}}             
\def\skytel{\aaref@jnl{S\&T}}             
\def\solphys{\aaref@jnl{Sol.~Phys.}}      
\def\sovast{\aaref@jnl{Soviet~Ast.}}      
\def\ssr{\aaref@jnl{Space~Sci.~Rev.}}     
\def\zap{\aaref@jnl{ZAp}}                 
\def\nat{\aaref@jnl{Nature}}              
\def\iaucirc{\aaref@jnl{IAU~Circ.}}       
\def\aplett{\aaref@jnl{Astrophys.~Lett.}} 
\def\apspr{\aaref@jnl{Astrophys.~Space~Phys.~Res.}}
\def\bain{\aaref@jnl{Bull.~Astron.~Inst.~Netherlands}} 
\def\fcp{\aaref@jnl{Fund.~Cosmic~Phys.}}  
\def\gca{\aaref@jnl{Geochim.~Cosmochim.~Acta}}   
\def\grl{\aaref@jnl{Geophys.~Res.~Lett.}} 
\def\jcp{\aaref@jnl{J.~Chem.~Phys.}}      
\def\jgr{\aaref@jnl{J.~Geophys.~Res.}}    
\def\jqsrt{\aaref@jnl{J.~Quant.~Spec.~Radiat.~Transf.}}
\def\memsai{\aaref@jnl{Mem.~Soc.~Astron.~Italiana}}
\def\nphysa{\aaref@jnl{Nucl.~Phys.~A}}   
\def\physrep{\aaref@jnl{Phys.~Rep.}}   
\def\physscr{\aaref@jnl{Phys.~Scr}}   
\def\planss{\aaref@jnl{Planet.~Space~Sci.}}   
\def\procspie{\aaref@jnl{Proc.~SPIE}}   

%% file: paper.bbl
\providecommand{\href}[2]{#2}\begingroup\raggedright\begin{thebibliography}{10}

\bibitem{Abbott:2017D}
{\bfseries Virgo, LIGO Scientific} Collaboration, ``{GW170817: Observation of
  Gravitational Waves from a Binary Neutron Star Inspiral},'' {\em Phys. Rev.
  Lett.} {\bfseries 119} (2017) 161101,
\href{http://arxiv.org/abs/1710.05832}{{\ttfamily arXiv:1710.05832 [gr-qc]}}.

\bibitem{Abbott:2020uma}
{\bfseries LIGO Scientific, Virgo} Collaboration, B.~Abbott {\em et~al.},
  ``{GW190425: Observation of a Compact Binary Coalescence with Total Mass
  $\sim 3.4 M_{\odot}$},''
  \href{http://dx.doi.org/10.3847/2041-8213/ab75f5}{{\em Astrophys. J. Lett.}
  {\bfseries 892} no.~1, (2020) L3},
  \href{http://arxiv.org/abs/2001.01761}{{\ttfamily arXiv:2001.01761
  [astro-ph.HE]}}.

\bibitem{Miller:2019cac}
M.~Miller {\em et~al.}, ``{PSR J0030+0451 Mass and Radius from $NICER$ Data and
  Implications for the Properties of Neutron Star Matter},''
  \href{http://dx.doi.org/10.3847/2041-8213/ab50c5}{{\em Astrophys. J. Lett.}
  {\bfseries 887} no.~1, (2019) L24},
  \href{http://arxiv.org/abs/1912.05705}{{\ttfamily arXiv:1912.05705
  [astro-ph.HE]}}.

\bibitem{Riley:2019yda}
T.~E. Riley {\em et~al.}, ``{A $NICER$ View of PSR J0030+0451: Millisecond
  Pulsar Parameter Estimation},''
  \href{http://dx.doi.org/10.3847/2041-8213/ab481c}{{\em Astrophys. J. Lett.}
  {\bfseries 887} no.~1, (2019) L21},
  \href{http://arxiv.org/abs/1912.05702}{{\ttfamily arXiv:1912.05702
  [astro-ph.HE]}}.

\bibitem{Bogdanov_2019}
S.~Bogdanov, F.~K. Lamb, S.~Mahmoodifar, M.~C. Miller, S.~M. Morsink, T.~E.
  Riley, T.~E. Strohmayer, A.~K. Tung, A.~L. Watts, A.~J. Dittmann,
  D.~Chakrabarty, S.~Guillot, Z.~Arzoumanian, and K.~C. Gendreau,
  ``Constraining the neutron star mass{\textendash}radius relation and dense
  matter equation of state with {NICER}. {II}. emission from hot spots on a
  rapidly rotating neutron star,''
  \href{http://dx.doi.org/10.3847/2041-8213/ab5968}{{\em The Astrophysical
  Journal} {\bfseries 887} no.~1, (Dec, 2019) L26}.
  \url{https://doi.org/10.3847%2F2041-8213%2Fab5968}.

\bibitem{Bilous_2019}
A.~V. Bilous, A.~L. Watts, A.~K. Harding, T.~E. Riley, Z.~Arzoumanian,
  S.~Bogdanov, K.~C. Gendreau, P.~S. Ray, S.~Guillot, W.~C.~G. Ho, and
  D.~Chakrabarty, ``A {NICER} view of {PSR} j0030+0451: Evidence for a
  global-scale multipolar magnetic field,''
  \href{http://dx.doi.org/10.3847/2041-8213/ab53e7}{{\em The Astrophysical
  Journal} {\bfseries 887} no.~1, (Dec, 2019) L23}.
  \url{https://doi.org/10.3847%2F2041-8213%2Fab53e7}.

\bibitem{Radice:2016rys}
D.~Radice, S.~Bernuzzi, W.~Del~Pozzo, L.~F. Roberts, and C.~D. Ott, ``{Probing
  Extreme-Density Matter with Gravitational Wave Observations of Binary Neutron
  Star Merger Remnants},''
  \href{http://dx.doi.org/10.3847/2041-8213/aa775f}{{\em Astrophys. J.}
  {\bfseries 842} no.~2, (2017) L10},
\href{http://arxiv.org/abs/1612.06429}{{\ttfamily arXiv:1612.06429
  [astro-ph.HE]}}.

\bibitem{Bauswein:2018bma}
A.~Bauswein, N.-U.~F. Bastian, D.~B. Blaschke, K.~Chatziioannou, J.~A. Clark,
  T.~Fischer, and M.~Oertel, ``{Identifying a first-order phase transition in
  neutron star mergers through gravitational waves},''
  \href{http://dx.doi.org/10.1103/PhysRevLett.122.061102}{{\em Phys. Rev.
  Lett.} {\bfseries 122} no.~6, (2019) 061102},
\href{http://arxiv.org/abs/1809.01116}{{\ttfamily arXiv:1809.01116
  [astro-ph.HE]}}.

\bibitem{Bauswein:2019skm}
A.~Bauswein, N.-U. Friedrich~Bastian, D.~Blaschke, K.~Chatziioannou, J.~A.
  Clark, T.~Fischer, H.-T. Janka, O.~Just, M.~Oertel, and N.~Stergioulas,
  ``{Equation-of-state Constraints and the QCD Phase Transition in the Era of
  Gravitational-Wave Astronomy},''
  \href{http://dx.doi.org/10.1063/1.5117803}{{\em AIP Conf. Proc.} {\bfseries
  2127} no.~1, (2019) 020013},
\href{http://arxiv.org/abs/1904.01306}{{\ttfamily arXiv:1904.01306
  [astro-ph.HE]}}.

\bibitem{PhysRevD.102.123023}
S.~Blacker, N.-U.~F. Bastian, A.~Bauswein, D.~B. Blaschke, T.~Fischer,
  M.~Oertel, T.~Soultanis, and S.~Typel, ``Constraining the onset density of
  the hadron-quark phase transition with gravitational-wave observations,''
  \href{http://dx.doi.org/10.1103/PhysRevD.102.123023}{{\em Phys. Rev. D}
  {\bfseries 102} (Dec, 2020) 123023}.
  \url{https://link.aps.org/doi/10.1103/PhysRevD.102.123023}.

\bibitem{Most:2018eaw}
E.~R. Most, L.~J. Papenfort, V.~Dexheimer, M.~Hanauske, S.~Schramm,
  H.~St{\"o}cker, and L.~Rezzolla, ``{Signatures of quark-hadron phase
  transitions in general-relativistic neutron-star mergers},''
  \href{http://dx.doi.org/10.1103/PhysRevLett.122.061101}{{\em Phys. Rev.
  Lett.} {\bfseries 122} no.~6, (2019) 061101},
  \href{http://arxiv.org/abs/1807.03684}{{\ttfamily arXiv:1807.03684
  [astro-ph.HE]}}.

\bibitem{Most:2019onn}
E.~R. Most, L.~J. Papenfort, V.~Dexheimer, M.~Hanauske, H.~St{\"o}cker, and
  L.~Rezzolla, ``{On the Deconfinement Phase Transition in Neutron-Star
  Mergers},''
\href{http://arxiv.org/abs/1910.13893}{{\ttfamily arXiv:1910.13893
  [astro-ph.HE]}}.

\bibitem{Weih:2019xvw}
L.~R. Weih, M.~Hanauske, and L.~Rezzolla, ``{Postmerger Gravitational-Wave
  Signatures of Phase Transitions in Binary Mergers},''
  \href{http://dx.doi.org/10.1103/PhysRevLett.124.171103}{{\em Phys. Rev.
  Lett.} {\bfseries 124} no.~17, (2020) 171103},
  \href{http://arxiv.org/abs/1912.09340}{{\ttfamily arXiv:1912.09340 [gr-qc]}}.

\bibitem{Ecker:2019xrw}
C.~Ecker, M.~Järvinen, G.~Nijs, and W.~van~der Schee, ``{Gravitational waves
  from holographic neutron star mergers},''
  \href{http://dx.doi.org/10.1103/PhysRevD.101.103006}{{\em Phys. Rev. D}
  {\bfseries 101} no.~10, (2020) 103006},
  \href{http://arxiv.org/abs/1908.03213}{{\ttfamily arXiv:1908.03213
  [astro-ph.HE]}}.

\bibitem{Chen:2019rja}
H.-Y. Chen, P.~M. Chesler, and A.~Loeb, ``{Searching for exotic cores with
  binary neutron star inspirals},''
  \href{http://dx.doi.org/10.3847/2041-8213/ab830f}{{\em Astrophys. J. Lett.}
  {\bfseries 893} no.~1, (2020) L4},
  \href{http://arxiv.org/abs/1909.04096}{{\ttfamily arXiv:1909.04096
  [astro-ph.HE]}}.

\bibitem{Annala:2019puf}
E.~Annala, T.~Gorda, A.~Kurkela, J.~Nättilä, and A.~Vuorinen, ``{Quark-matter
  cores in neutron stars},'' \href{http://arxiv.org/abs/1903.09121}{{\ttfamily
  arXiv:1903.09121 [astro-ph.HE]}}.

\bibitem{Liebling:2020jlq}
S.~Liebling, C.~Palenzuela, and L.~Lehner, ``{Towards fidelity and scalability
  in non-vacuum mergers},'' \href{http://arxiv.org/abs/2002.07554}{{\ttfamily
  arXiv:2002.07554 [gr-qc]}}.

\bibitem{arbona13}
A.~{Arbona}, A.~{Artigues}, C.~{Bona-Casas}, J.~{Mass{\'o}}, B.~{Mi{\~n}ano},
  A.~{Rigo}, M.~{Trias}, and C.~{Bona}, ``{Simflowny: A general-purpose
  platform for the management of physical models and simulation problems},''
  \href{http://dx.doi.org/10.1016/j.cpc.2013.04.012}{{\em Computer Physics
  Communications} {\bfseries 184} (Oct., 2013) 2321--2331}.

\bibitem{arbona18}
A.~{Arbona}, B.~{Mi{\~n}ano}, A.~{Rigo}, C.~{Bona}, C.~{Palenzuela},
  A.~{Artigues}, C.~{Bona-Casas}, and J.~{Mass{\'o}}, ``{Simflowny 2: An
  upgraded platform for scientific modelling and simulation},''
  \href{http://dx.doi.org/10.1016/j.cpc.2018.03.015}{{\em Computer Physics
  Communications} {\bfseries 229} (Aug., 2018) 170--181},
  \href{http://arxiv.org/abs/1702.04715}{{\ttfamily arXiv:1702.04715 [cs.MS]}}.

\bibitem{hornung02}
R.~D. Hornung and S.~R. Kohn, ``Managing application complexity in the samrai
  object-oriented framework,'' \href{http://dx.doi.org/10.1002/cpe.652}{{\em
  Concurrency and Computation: Practice and Experience} {\bfseries 14} no.~5,
  (2002) 347--368}. \url{http://dx.doi.org/10.1002/cpe.652}.

\bibitem{gunney16}
B.~T. Gunney and R.~W. Anderson, ``Advances in patch-based adaptive mesh
  refinement scalability,''
  \href{http://dx.doi.org/https://doi.org/10.1016/j.jpdc.2015.11.005}{{\em
  Journal of Parallel and Distributed Computing} {\bfseries 89} (2016) 65 --
  84}.
  \url{http://www.sciencedirect.com/science/article/pii/S0743731515002129}.

\bibitem{Palenzuela:2018sly}
C.~Palenzuela, B.~Miñano, D.~Viganò, A.~Arbona, C.~Bona-Casas, A.~Rigo,
  M.~Bezares, C.~Bona, and J.~Massó, ``{A Simflowny-based finite-difference
  code for high-performance computing in numerical relativity},''
  \href{http://dx.doi.org/10.1088/1361-6382/aad7f6}{{\em Class. Quant. Grav.}
  {\bfseries 35} no.~18, (2018) 185007},
\href{http://arxiv.org/abs/1806.04182}{{\ttfamily arXiv:1806.04182
  [physics.comp-ph]}}.

\bibitem{2020PhRvD.101l3019V}
D.~{Vigan{\`o}}, R.~{Aguilera-Miret}, F.~{Carrasco}, B.~{Mi{\~n}ano}, and
  C.~{Palenzuela}, ``{General relativistic MHD large eddy simulations with
  gradient subgrid-scale model},''
  \href{http://dx.doi.org/10.1103/PhysRevD.101.123019}{{\em \prd} {\bfseries
  101} no.~12, (June, 2020) 123019},
  \href{http://arxiv.org/abs/2004.00870}{{\ttfamily arXiv:2004.00870 [gr-qc]}}.

\bibitem{Butcher:2008}
J.~C. Butcher, \href{http://dx.doi.org/10.1002/9780470753767.fmatter}{{\em
  Numerical Methods for Ordinary Differential Equations}}.
\newblock John Wiley and Sons, Ltd, 2008.
\newblock \url{http://dx.doi.org/10.1002/9780470753767.fmatter}.

\bibitem{toro97}
E.~F. Toro, {\em {Riemann solvers and numerical methods for fluid dynamics: a
  practical introduction; 2nd ed.}}
\newblock Springer, Berlin, 1999.
\newblock \url{http://cds.cern.ch/record/404378}.

\bibitem{shu98}
C.-W. Shu, {\em Essentially non-oscillatory and weighted essentially
  non-oscillatory schemes for hyperbolic conservation laws},
  \href{http://dx.doi.org/10.1007/BFb0096355}{pp.~325--432}.
\newblock Springer Berlin Heidelberg, Berlin, Heidelberg, 1998.
\newblock \url{https://doi.org/10.1007/BFb0096355}.

\bibitem{suresh97}
A.~Suresh and H.~Huynh, ``Accurate monotonicity-preserving schemes with
  runge–kutta time stepping,''
  \href{http://dx.doi.org/https://doi.org/10.1006/jcph.1997.5745}{{\em Journal
  of Computational Physics} {\bfseries 136} no.~1, (1997) 83 -- 99}.
  \url{http://www.sciencedirect.com/science/article/pii/S0021999197957454}.

\bibitem{Abbott:2018exr}
{\bfseries LIGO Scientific, Virgo} Collaboration, B.~Abbott {\em et~al.},
  ``{GW170817: Measurements of neutron star radii and equation of state},''
  \href{http://dx.doi.org/10.1103/PhysRevLett.121.161101}{{\em Phys. Rev.
  Lett.} {\bfseries 121} no.~16, (2018) 161101},
  \href{http://arxiv.org/abs/1805.11581}{{\ttfamily arXiv:1805.11581 [gr-qc]}}.

\bibitem{Lindblom:1998dp}
L.~Lindblom, ``{Phase transitions and the mass radius curves of relativistic
  stars},'' \href{http://dx.doi.org/10.1103/PhysRevD.58.024008}{{\em Phys.
  Rev.} {\bfseries D58} (1998) 024008},
\href{http://arxiv.org/abs/gr-qc/9802072}{{\ttfamily arXiv:gr-qc/9802072
  [gr-qc]}}.

\bibitem{lorene}
``{\sc Lorene} home page.'' \url{http://www.lorene.obspm.fr/}, 2010.

\bibitem{Ferdman:2020huz}
R.~Ferdman {\em et~al.}, ``{Asymmetric mass ratios for bright double
  neutron-star mergers},''
  \href{http://dx.doi.org/10.1038/s41586-020-2439-x}{{\em Nature} {\bfseries
  583} no.~7815, (2020) 211--214},
  \href{http://arxiv.org/abs/2007.04175}{{\ttfamily arXiv:2007.04175
  [astro-ph.HE]}}.

\bibitem{PhysRevLett.76.4878}
D.~Lai, ``Tidal stablization of neutron stars and white dwarfs,''
  \href{http://dx.doi.org/10.1103/PhysRevLett.76.4878}{{\em Phys. Rev. Lett.}
  {\bfseries 76} (Jun, 1996) 4878--4881}.
  \url{https://link.aps.org/doi/10.1103/PhysRevLett.76.4878}.

\bibitem{2010PhRvD..81h4016D}
T.~{Damour} and A.~{Nagar}, ``{Effective one body description of tidal effects
  in inspiralling compact binaries},''
  \href{http://dx.doi.org/10.1103/PhysRevD.81.084016}{{\em PRD} {\bfseries 81}
  no.~8, (Apr., 2010) 084016}, \href{http://arxiv.org/abs/0911.5041}{{\ttfamily
  arXiv:0911.5041 [gr-qc]}}.

\bibitem{PhysRevLett.112.201101}
S.~Bernuzzi, A.~Nagar, S.~Balmelli, T.~Dietrich, and M.~Ujevic,
  ``Quasiuniversal properties of neutron star mergers,''
  \href{http://dx.doi.org/10.1103/PhysRevLett.112.201101}{{\em Phys. Rev.
  Lett.} {\bfseries 112} (May, 2014) 201101}.
  \url{https://link.aps.org/doi/10.1103/PhysRevLett.112.201101}.

\bibitem{Lehner:2016lxy}
L.~Lehner, S.~L. Liebling, C.~Palenzuela, O.~Caballero, E.~O'Connor,
  M.~Anderson, and D.~Neilsen, ``{Unequal mass binary neutron star mergers and
  multimessenger signals},''
  \href{http://dx.doi.org/10.1088/0264-9381/33/18/184002}{{\em Class. Quant.
  Grav.} {\bfseries 33} no.~18, (2016) 184002},
  \href{http://arxiv.org/abs/1603.00501}{{\ttfamily arXiv:1603.00501 [gr-qc]}}.

\bibitem{Vretinaris:2019spn}
S.~Vretinaris, N.~Stergioulas, and A.~Bauswein, ``{Empirical relations for
  gravitational-wave asteroseismology of binary neutron star mergers},''
  \href{http://dx.doi.org/10.1103/PhysRevD.101.084039}{{\em Phys. Rev. D}
  {\bfseries 101} no.~8, (2020) 084039},
  \href{http://arxiv.org/abs/1910.10856}{{\ttfamily arXiv:1910.10856 [gr-qc]}}.

\bibitem{2016PhRvD..93l4051R}
L.~{Rezzolla} and K.~{Takami}, ``{Gravitational-wave signal from binary neutron
  stars: A systematic analysis of the spectral properties},''
  \href{http://dx.doi.org/10.1103/PhysRevD.93.124051}{{\em \prd} {\bfseries 93}
  no.~12, (June, 2016) 124051},
  \href{http://arxiv.org/abs/1604.00246}{{\ttfamily arXiv:1604.00246 [gr-qc]}}.

\bibitem{Paschalidis:2015mla}
V.~Paschalidis, W.~E. East, F.~Pretorius, and S.~L. Shapiro, ``{One-arm Spiral
  Instability in Hypermassive Neutron Stars Formed by Dynamical-Capture Binary
  Neutron Star Mergers},''
\href{http://arxiv.org/abs/1510.03432}{{\ttfamily arXiv:1510.03432
  [astro-ph.HE]}}.

\bibitem{Lehner:2016wjg}
L.~Lehner, S.~L. Liebling, C.~Palenzuela, and P.~M. Motl, ``{m=1 instability
  and gravitational wave signal in binary neutron star mergers},''
  \href{http://dx.doi.org/10.1103/PhysRevD.94.043003}{{\em Phys. Rev.}
  {\bfseries D94} no.~4, (2016) 043003},
\href{http://arxiv.org/abs/1605.02369}{{\ttfamily arXiv:1605.02369 [gr-qc]}}.

\bibitem{Yang:2018bzx}
H.~Yang, W.~E. East, V.~Paschalidis, F.~Pretorius, and R.~F. Mendes,
  ``{Evolution of Highly Eccentric Binary Neutron Stars Including Tidal
  Effects},'' \href{http://dx.doi.org/10.1103/PhysRevD.98.044007}{{\em Phys.
  Rev. D} {\bfseries 98} no.~4, (2018) 044007},
  \href{http://arxiv.org/abs/1806.00158}{{\ttfamily arXiv:1806.00158 [gr-qc]}}.

\bibitem{Yang:2017xlf}
H.~Yang, V.~Paschalidis, K.~Yagi, L.~Lehner, F.~Pretorius, and N.~Yunes,
  ``{Gravitational wave spectroscopy of binary neutron star merger remnants
  with mode stacking},''
  \href{http://dx.doi.org/10.1103/PhysRevD.97.024049}{{\em Phys. Rev. D}
  {\bfseries 97} no.~2, (2018) 024049},
  \href{http://arxiv.org/abs/1707.00207}{{\ttfamily arXiv:1707.00207 [gr-qc]}}.

\bibitem{Whitaker}
T.~Whitaker, W.~E. East, S.~Green, L.~Lehner, and H.~Yang, ``Phenomenological
  waveforms for post-merger binary neutron stars with artificial neural
  networks.'' In preparation.

\end{thebibliography}\endgroup
